\documentclass[journal,twoside,web]{ieeecolor}
\usepackage{cite}
\usepackage{amsmath,amssymb,amsfonts}
\usepackage{algorithmic}
\usepackage{graphicx}
\usepackage{textcomp}
\usepackage{booktabs}
\usepackage{multirow}
\usepackage{generic}
\usepackage{algorithm,algorithmic}
\usepackage{hyperref}
\hypersetup{hidelinks=true}
\def\BibTeX{{\rm B\kern-.05em{\sc i\kern-.025em b}\kern-.08em
    T\kern-.1667em\lower.7ex\hbox{E}\kern-.125emX}}
\begin{document}
\title{Re-Visible Dual-Domain Self-Supervised Deep Unfolding Network for MRI Reconstruction }
\author{Hao Zhang, Qi Wang, Jian Sun, Zhijie Wen, Jun Shi and Shihui Ying, \IEEEmembership{Member, IEEE}
\thanks{This work is supported by the National Key R \& D Program of
China under Grant 2021YFA1003004, the National Natural Science Foundation of China (Grant No.12571566), the openFund for Key Laboratory of Internet of Aircrafts (No.MHFLW202305) and the Large-scale Numerical Simulation Computing Sharing Platform of Shanghai University. (Corresponding author: Zhijie Wen.) }
\thanks{Hao Zhang, Qi Wang, Zhijie Wen are with the Department of Mathematics, School of Science, Shanghai University, Shanghai 200444, China. (e-mail: Zhanghao123@shu.edu.cn; wangqi2020@shu.edu.cn; wenzhijie@shu.edu.cn).}
\thanks{Shihui Ying is with Shanghai Institute of Applied Mathematics and Mechanics and the School of Mechanics and Engineering Science, Shanghai University, Shanghai 200072, China. (e-mail: shying@shu.edu.cn).}
\thanks{Jian Sun is with the School of Mathematics and Statistics,
Xi’an Jiaotong University, Xi’an, Shaanxi, 710049, P. R. China. (e-mail: jiansun@xjtu.edu.cn).}
\thanks{Jun Shi is with the School of Communication and Information Engineering, Shanghai University, Shanghai 200444, China. (e-mail:  junshi@shu.edu.cn).}}

\maketitle

\begin{abstract}
Magnetic Resonance Imaging (MRI) is widely used in clinical practice, but {suffers} from prolonged acquisition time. Although deep learning methods have been proposed to accelerate acquisition and demonstrate promising performance, they rely on high-quality fully-sampled datasets for training in a supervised manner. However, such datasets are time-consuming and expensive-to-collect, which constrains their broader applications. On the other hand, self-supervised methods offer an alternative by enabling learning from under-sampled data alone, but most existing methods rely on further partitioned under-sampled k-space data as model's input for training, which causes {an input distribution shift between the the training stage and the inference stage.} Additionally, their models have not effectively incorporated comprehensive image priors, leading to degraded reconstruction performance. In this paper, we propose a novel re-visible dual-domain self-supervised deep unfolding network to address these issues when only under-sampled datasets are available. Specifically, by incorporating re-visible dual-domain loss, all under-sampled k-space data are utilized during training to mitigate {the input distribution shift} caused by further partitioning. This design enables the model to implicitly adapt to all under-sampled k-space data as input. Additionally, we design a Deep Unfolding Network based on Chambolle and Pock Proximal Point Algorithm (DUN-CP-PPA) to achieve end-to-end reconstruction. {By employing a Spatial-Frequency Feature Extraction (SFFE) block to capture both global and local representations, the model effectively integrates imaging physics with comprehensive image priors to enhance reconstruction performance. Experiments on both single-coil and multi-coil datasets demonstrate that our method outperforms state-of-the-art approaches in terms of reconstruction performance and generalization capability.}
\end{abstract}


\section{Introduction}
\label{sec:introduction}
\IEEEPARstart{M}{agnetic} Resonance Imaging (MRI) is a widely employed medical imaging technique, known for its non-invasive approach, high resolution, and superior soft tissue contrast \cite{b37}. Besides, the integration of various imaging modalities provides additional insights into tissue and organ characteristics, improving diagnostic precision. However, acquisition of certain MRI modalities {tends} to be slow because of the repetitive signal spatial encoding and the constraints of the hardware. The prolonged acquisition time increases {patients' discomfort} and leads to a higher accumulation of motion artifacts, which can degrade image quality and compromise diagnostic accuracy. Consequently, accelerating MRI acquisition is crucial for clinical practice. One effective approach \cite{b1} to achieve this is by reducing the amount of k-space data collected by a pre-defined pattern, and subsequently reconstructing fully-sampled images from under-sampled data.

{An effective strategy for accelerating MRI acquisition is to exploit the inherent redundancy among signals acquired by multiple coils in Parallel Imaging (PI). However, such methods often suffer from residual aliasing artifacts under high acceleration factors. To address this limitation, Compressed Sensing Magnetic Resonance Imaging (CS-MRI) techniques \cite{b2,b3,b4,b50} enable accurate image reconstruction from under-sampled data at sampling rates far below those dictated by the Nyquist sampling theorem. Methods integrating the strengths of PI and CS-MRI typically leverage the sparsity of signals and optimization algorithms to further reduce the sampling requirements while preserving high image quality.} Although theoretically sound, crafting an optimal regularizer remains a difficult task. Recent deep learning-based approaches \cite{b5,b6,b19,b38,b39,b40,b47,b48,b49} have shown promising results, enabling fast and accurate reconstructions. However, these methods are often "black-boxes," lacking interpretability and physical insight, which limits their clinical applications. To overcome this limitation, deep unfolding networks \cite{b7,b8,b9,b10,b11,b12,b20,b24,b41,b42} have been introduced. These networks combine the imaging physics and image priors by effectively unfolding the iterations of an optimization algorithm into deep neural networks, which enhances both interpretability and performance. Although deep unfolding networks have shown promise in MRI reconstruction, most approaches rely on supervised training on high-quality fully-sampled datasets. However, in clinical practice, collecting such datasets is both time-consuming and costly, may even be technically unfeasible, as it requires extended acquisition times, advanced equipment, and specialized expertise. For example, cardiovascular MRI is challenged by excessive involuntary movements, while diffusion MRI with echo-planar imaging suffers from rapid $T^{*}_2$ signal decay.

To address this limitation, self-supervised learning \cite{b13,b14,b15,b16,b17,b18,b21,b22,b23,b32,b45,b46}, which enables learning from only under-sampled k-space data, has gained significant interest from researchers. For example, Huang et al. \cite{b13} replace the under-sampled and fully-sampled training pairs with pairs of under-sampled data when training the network, reducing the dependence of model training on high-quality datasets to some extent. However, obtaining pairs of under-sampled data is still quite challenging in practice. Some methods \cite{b14,b15,b16,b18,b22,b23} have been developed to enable training using only single under-sampled data. For instance, Yaman et al. \cite{b23} {divide} the acquired k-space data into two disjoint sets, one used as the input and the other as the target for cross-validation. Although the dependence on high-quality datasets has been significantly reduced, most existing methods rely on further partitioned under-sampled k-space data as input for training, which causes {an input distribution shift between
the the training stage and the inference stage.} Additionally, their models have not effectively incorporated comprehensive image priors, leading to degraded reconstruction performance.

In this paper, we propose a re-visible dual-domain self-supervised deep unfolding network to address the aforementioned limitations. Our framework integrates a re-visible dual-domain self-supervised learning approach to mitigate {the input distribution shift between the the training stage and the inference stage} when only under-sampled datasets are available. Furthermore, it incorporates a Deep Unfolding Network based on the Chambolle and Pock Proximal Point Algorithm (DUN-CP-PPA), embedding the imaging model directly into the reconstruction process to improve incorporation of comprehensive image priors. Specifically, by introducing re-visible dual-domain loss, all under-sampled k-space data are utilized during training, effectively mitigate {the input distribution shift} caused by further partitioning. This design allows the model to implicitly adapt to all under-sampled k-space data as input (make the partitioned k-space data "re-visible" to the model). {Moreover, we adopt DUN-CP-PPA with an integrated Spatial–Frequency Feature Extraction (SFFE) block that jointly captures global and local representations, enabling end-to-end MRI reconstruction through optimization unfolding. This design effectively combines imaging physics with comprehensive image priors, maintaining network transparency and yielding superior reconstruction quality.}
The main contributions of this paper are listed as follows:

\begin{itemize}
    \item We propose a re-visible dual-domain self-supervised deep unfolding network to improve reconstruction performance utilizing only under-sampled k-space data for training.
    \item We propose a re-visible dual-domain self-supervised learning framework that incorporates a re-visible dual-domain loss, {enabling the model to implicitly adapt to all under-sampled k-space data and thereby mitigating the input distribution shift caused by further partitioning.}
    \item By unfolding each stage of CP-PPA with network modules, including SFFE block that jointly captures global and local representations, we embed imaging physics and comnprehensive image priors to guide the reconstruction process, resulting in DUN-CP-PPA. 
    \item Through extensive experiments on both the single-coil and multi-coil datasets, we show that the proposed model outperforms current state-of-the-art methods reconstruction performance and generalization capability.
\end{itemize}

\section{Related work}
\subsection{Chambolle and Pock Proximal Point Algorithm}
\label{cpppa}
The Chambolle and Pock Proximal Point Algorithm (CP-PPA) is an effective optimization technique \cite{b26,b27,b28} that employs a primal-dual hybrid approach \cite{b29} to ensure both convergence and efficiency. The original objective function:

\[
\min_x f(x) + g(L x), \tag{1}
\]
is transformed into its primal-dual form:

\[
\min_x \max_y f(x) + \langle L x, y \rangle - g^*(y),\tag{2}
\]
where \( g^* \) is the convex conjugate of \( g \), \( L \) is a continuous linear operator . {The Karush–Kuhn–Tucker (KKT) conditions} for this problem are:

\[
\left\{
\begin{array}{l}
-L^H y_{\star} \in \partial f(x_{\star}), \\
L x_{\star} \in \partial g^*(y_{\star}),
\end{array} 
\right. \tag{3}
\]
where \(L^H\) is the Hermitian transpose of \(L\). CP-PPA solves the primal-dual problem iteratively using the update rules:

\[
\left\{
\begin{aligned}
x_{k+1} &= \operatorname{Prox}_f^{\tau_{k+1}} \left( x_k - \tau_{k+1} L^H y_k \right), \\
y_{k+1} &= \operatorname{Prox}_{g^*}^{\sigma_{k+1}} ( y_k + \sigma_{k+1} L z_{k+1}).
\end{aligned}\tag{4}
\right.
\]
where $z_{k+1} = x_{k+1} + \theta_{k+1} ( x_{k+1} - x_k )$. {$\tau_{k}$ and $\sigma_{k}$ are step-size parameters, and $\operatorname{Prox}_{h}^{\lambda}(x)=\arg\min_y \left\{h(y)+\frac{1}{2\lambda}\|y-x\|^2\right\}$ denotes the proximal mapping associated with a function $h$ and a step-size parameter $\lambda$. }
Compared to {the Iterative Shrinkage-Thresholding Algorithm (ISTA)} \cite{b30}, CP-PPA converges faster and provides stronger theoretical guarantees for globally optimal solutions \cite{b28}. {Compared to ADMM, replacing the complex linear least-squares subproblem with an explicit proximal operator update greatly reduces computational complexity and leads to more efficient iterations.
Benefiting from its convergence speed and convergence guarantees, the CP-PPA framework has been widely adopted to solve various inverse problems, including image denoising \cite{b57}, magnetic resonance imaging (MRI) reconstruction \cite{b56}, and computed tomography (CT) reconstruction \cite{b58}. 

In the context of MRI reconstruction, the optimization formulation naturally fits the problem structure. The forward operator $L$ is defined as the composition of the Fourier transform $\mathcal{F}$ and the sampling mask $M$, i.e., $L=F_m=M \mathcal{F}$. The reconstruction task is typically formulated as minimizing the sum of a regularization term $f(x)$, which encodes prior knowledge such as sparsity, and a data fidelity term $g\left(F_m x\right)=\left\|F_m x-\tilde{k}\right\|_1$, which enforces consistency between the reconstructed image $x$ and the acquired under-sampled k-space data $\tilde{k}$. Recent studies \cite{b24,b59,b60} have demonstrated the effectiveness and potential of CP-PPA in accelerating MRI reconstruction, achieving improved image quality.}
\begin{figure*}[!t]
\centerline{\includegraphics[width=0.9\textwidth]{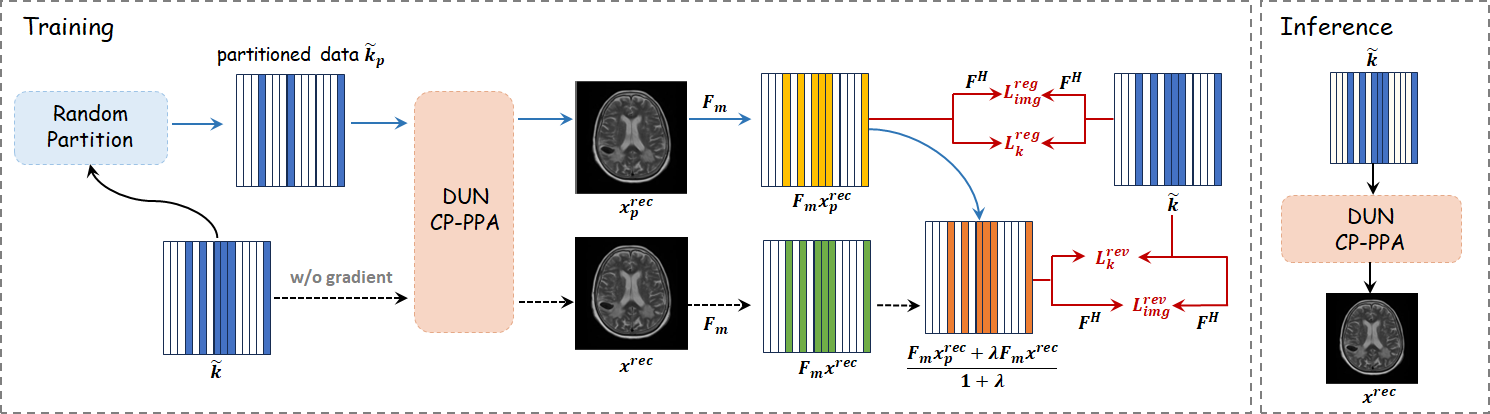}}
\caption{The framework of our proposed re-visible dual-domain self-supervised learning is shown. We utilize DUN-CP-PPA for reconstruction with image physics incorporated. During training, we adopt a dual-branch structure. The first branch reconstructs \( \tilde{k}_p \) to boost the model’s generative capability and the second branch reconstructs \( \tilde{k} \) to mitigate {the input distribution shift} caused by further partitioning. The model is trained using the proposed re-visible dual-domain loss. During inference, the trained network can directly reconstruct the image from \( \tilde{k} \).
}
\label{fig2}
\end{figure*}

\subsection{Deep Unfolding Network}
Deep unfolding networks effectively combine model-based optimization algorithms with data-driven learning approaches, providing a combination of the advantages of both approaches. They have attracted significant attention in recent years due to their impressive performance, especially in CS-MRI \cite{b7,b8,b9,b10,b11,b12}.

In CS-MRI, Deep Unfolding Networks have demonstrated considerable success by unfolding various optimization techniques. ADMM-CSNet \cite{b10,b12,b13} casts the iterative Alternating Direction Method of Multipliers (ADMM) algorithm into a deep network architecture for image CS reconstruction. Zhang et al. \cite{b62} and Zhang et al. \cite{b64} integrate the Alternating Iterative Shrinkage-Thresholding Algorithm (ISTA) into the model design. Several studies \cite{b7,b8,b9} expand Half-Quadratic Splitting (HQS) with corresponding network modules. Different network modules also enhance the capabilities of deep unfolding networks, such as CNNs and U-Nets.  Liang et al. \cite{b25} {propose} an Adaptive Local Neighborhood-based Neural Network that integrates with deep unfolding networks to improve reconstruction quality. Yang et al. \cite{b11} {enhance} the network's ability to extract comprehensive features by incorporating channel and spatial attention modules. Sun et al. \cite{b49} and Luoet al. \cite{b51} employ Vision Transformers to capture global features, thereby enhancing the overall feature representation. Integrating both global and local features offers a more comprehensive strategy. Lei et al. \cite{b52} and Jiang et al.\cite{b12} {introduce} a dual-domain feature fusion approach, where local features extracted in the spatial domain are fused with global features extracted in the frequency domain, both obtained through convolution operations. 

{Different from existing method, we directly modulates all frequency components through the learnable matrix, enabling the model to capture long-range spatial dependencies in the frequency domain with log-linear complexity. }

\subsection{Self-Supervised Learning Methods}
In order to get rid of dependence on fully-sampled data, self-supervised methods \cite{b13,b14,b15,b16,b17,b18,b21,b22,b23,b32} are proposed for MRI reconstruction, which are trained with under-sampled data. The classical k-space interpolation method \cite{b33,b34}, which first fully samples a calibrated region in the central part of the k-space, learns a linear kernel and uses it in a translation-invariant manner to interpolate in the missing k-space data. Inspired by deep learning, linear kernels are generalized to convolutional neural networks \cite{b35,b36} to improve the accuracy of missing data interpolation. With the further development of self-supervised learning techniques, many works have also emerged in the field of MRI reconstruction. By further down-sampling, a new self-supervised learning framework is designed \cite{b15,b16,b22,b23,b32,b13,b18}, requiring only under-sampled k-space data itself. Huang et al.\cite{b13} uses pairs of under-sampled k-space data for training, which reduces the difficulty of data collection. Klug et al. \cite{b15} and Yaman et al. \cite{b23} {divide} the k-space data into two disjoint sets, using one as input and the other as supervision targets. Yan et al. \cite{b16} and Hu et al. \cite{b32} propose to use a parallel training framework for self-supervised MRI reconstruction. Zhou et al. \cite{b18} {design} a triple-branch-based dual-domain self-supervised reconstruction framework, achieving promising performance on single-contrast MRI reconstruction. Zhou et al. \cite{b14} {combine} the Swin Transformer to leverage non-local processing to recover fine details. To learn the high-quality universal representation of MR images, an attention-weighted average pooling module \cite{b16} is used in conjunction with a contrastive learning strategy. 

{In contrast to prior works, we focus on mitigating the input distribution shift caused by further partitioning between the the training stage and the inference stage, where the latter operates on fully under-sampled data.}

\section{Method}
{
\subsection{Model Overview}
Deep learning-based methods have been widely applied to accelerate MRI reconstruction. However, their performance heavily depends on the quantity and quality of training data. Acquiring paired or fully sampled data is often costly and time-consuming. Moreover, many existing models have not effectively incorporated comprehensive image priors, resulting in degraded reconstruction performance.

To mitigate the data dependency issue, some studies adopt self-supervised learning strategies that use further partitioned and fully under-sampled k-space data as training pairs. In this approach, the network, without explicit knowledge of which k-space entries are additionally masked, is trained to minimize the reconstruction loss over the entire k-space. As a result, the network learns to map under-sampled k-space data to fully sampled MR images. Empirically, using fully under-sampled k-space data during inference demonstrates better performance. However, using only partially under-sampled k-space data as input during training inevitably leads to an input distribution shift between the the training stage and the inference stage, while directly using fully under-sampled data for training may cause the network to degenerate into an identity mapping. To address this issue, we propose using further partitioned k-space data as an intermediate, allowing the fully under-sampled k-space data to be implicitly involved in the training process. The remaining challenge lies in designing an effective framework to realize this idea.

To address the second limitation, inspired by the advantages of CP-PPA discussed in Section \ref{cpppa}, we propose DUN-CP-PPA, a deep unfolding network built upon the CP-PPA optimization framework. This model effectively integrates the physical model of MRI acquisition into the unfolding architecture. Each layer of the DUN corresponds to the iterations of a physics-informed optimization process, ensuring data consistency, while the embedded network modules perform adaptive regularization by extracting both local feature representations from the spatial domain and global feature representations from the frequency domain.
}

Our framework is shown in Fig. \ref{fig2}. We adopt a dual-branch structure for self-supervised learning and utilize DUN-CP-PPA for reconstruction. It consists of two key components: (1) a re-visible dual-domain self-supervised learning approach, introduced in \ref{DDSL}, {which enables learning only from under-sampled data and mitigate the input distribution shift caused by further partitioning}; and (2) a DUN-CP-PPA, 
{discussed in \ref{DUN-CP-PPA}, built upon the CP-PPA optimization framework, which effectively integrates the physical model of MRI acquisition and jointly incorporates comprehensive image priors across spatial and frequency domains to improve reconstruction performance.}

\subsection{Re-Visible Dual-Domain Self-Supervised Learning}
\label{DDSL}
\begin{figure*}[!t]
\centerline{\includegraphics[width=0.9\textwidth]{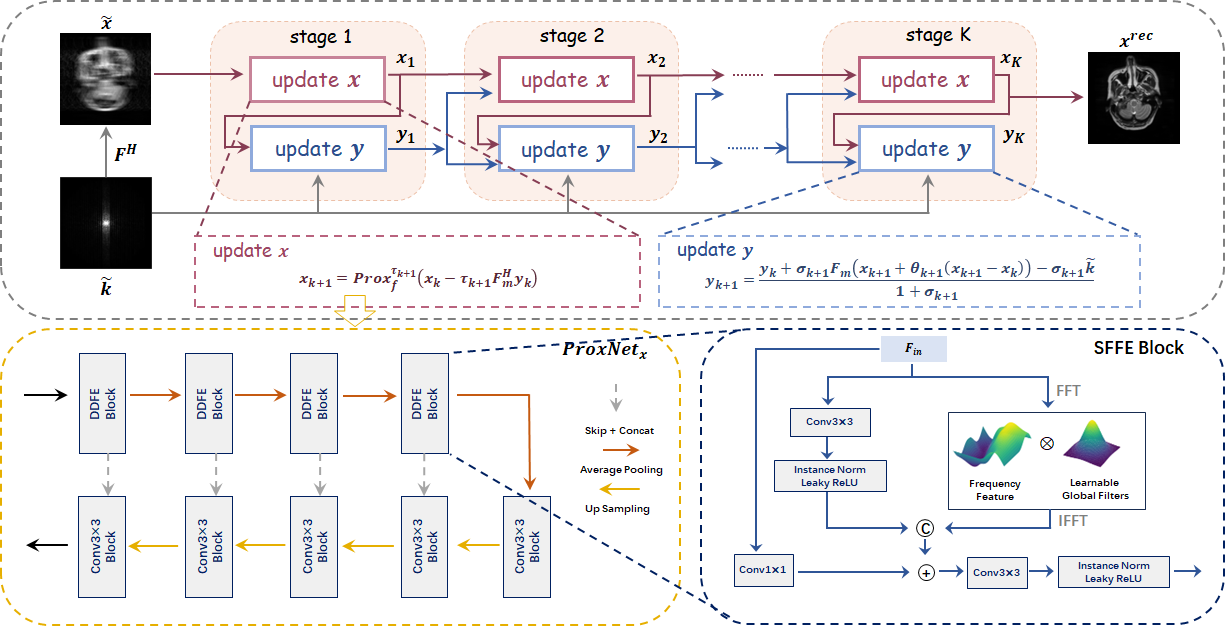}}
\caption{The overall structure of the proposed Chambolle and Pock Proximal Point Algorithm based Deep Unfolding Network (DUN-CP-PPA) is shown in the first line, which includes two parts: update \( x \) module and update \( y \) module. Among them, the update \( y \) module is in an analytical form and is computed according to the given equation. The update \( x \) module includes \( \operatorname{ProxNet}_x \), which is shown in the second line, and extracts multi-scale features using the U-Net architecture. In \( \operatorname{ProxNet}_x \), there is a Spatial-Frequency Feature Extraction Block that captures comprehensive feature representations.
}
\label{fig1}
\end{figure*}
{The concept of “re-visible” originates from the process visible → invisible → visible. Initially, the under-sampled k-space data are visible and can be directly used for both training and inference. As adopted in previous work \cite{b23}, we introduce further under-sampling operation which makes part of the data invisible to the model, resulting in {an input distribution shift} between the the training stage and the inference stage, which in turn limits the model’s generalization ability. To make these crucial yet invisible data visible to the model again, we propose a re-visible self-supervised learning framework. Specifically, the invisible part is implicitly incorporated into training through a specially designed secondary branch, enabling the model to adapt to them during learning. More detailed discussion about incorporating \( \tilde{k} \) during training is presented in Appendix \ref{A2}.}

As discussed above, we propose a re-visible dual-domain self-supervised learning approach as illustrated in Fig. \ref{fig2}. We first perform a random partition on the under-sampled k-space data \( \tilde{k} \) to obtain \( \tilde{k}_p \). Then, \( \tilde{k} \) and \( \tilde{k}_p \) are fed into the DUN-CP-PPA to obtain the corresponding reconstructed images \( x^{rec} \) and \( x^{rec}_p \). During training, we incorporate a re-visible dual-domain loss to {enable the fully under-sampled k-space data to participate in training, mitigating the input distribution shift between the the training stage and the inference stage when only under-sampled datasets are available.}

In k-space, the first branch reconstructs \( \tilde{k}_p \), which is supervised by \( \tilde{k} \) through the corresponding loss function \(\|F_m x^{rec}_p - \tilde{k}\|_1\), {where \( F_m x^{rec}_p \) denotes the Fourier transform of image \( x^{rec}_p \) followed by the application of a sampling mask,} ensuring the model's generative capability is effectively trained. However, after partitioning, \( \tilde{k}_p \) becomes {partially available} and leads to {an input distribution shift} between the the training stage and the inference stage. To address this issue, the second branch reconstructs \( \tilde{k} \), ensuring that all under-sampled k-space data contribute to the training process. Inspired by \cite{b43}, it is worth noting that directly utilizing \(\mathcal{L} = \|F_mx^{rec} - \tilde{k}\|_1 + \lambda\|F_mx_p^{rec} - \tilde{k}\|_1\) as the loss function may learn the identity and degrade the reconstruction performance, because \( x^{rec} \) is explicitly involved in backpropagation. In order to fully utilize all k-space data while avoiding learning identity, we hope $x_{rec}$ to implicitly participate in backpropagation, allowing the model to adapt to it during the training phase. Therefore, we use the following loss function for the second branch{, obtaining the re-visible term in k-space \( L_k^{rev} \).}
\[
\mathcal{L}_{k}^{rev} = {\left\|\frac{F_m x^{rec}_p + \lambda F_m sg(x^{rec})}{1 + \lambda} - \tilde{k}\right\|_1}. \tag{5} \label{eq11}
\]

In this formulation, \( \text{sg}(\cdot) \) represents the stop-gradient operation, which prevents \( x^{rec} \) from explicitly participating in backpropagation to avoid identity mapping issue. The parameter \( \lambda \) is a hyperparameter that controls usage intensity of all k-space {data}. Through this setup, \( x^{rec} \) implicitly participates in training, ensuring that the model can access all under-sampled k-space data during the training process. Specifically, by incorporating this re-visible term, \( x_p^{rec} \) is used as a transition to optimize \( x^{rec} \), for the reason that \(\mathcal{L}_{k}^{rev}\) is affected by reconstruction performance of \(x^{rec}\). Consequently, the model facilitates an implicit adaptation process, from optimizing the reconstruction result \( x_p^{rec} \), which is derived from partially under-sampled k-space data, to implicitly optimizing the reconstruction result \( x^{rec} \), which is derived from all under-sampled k-space data in order to further reduce the loss. This adaptation enables the model to implicitly learn to utilize complete k-space data as input.

We combine dual-branch efforts, using \( \mathcal{L}_{k}^{rev} \) to involve all under-sampled k-space data during training, thereby mitigating {the input distribution shift between the the training stage and the inference stage}. Additionally, we employ \(\|F_m x^{rec}_p - \tilde{k}\|_1\) as a regularization term to ensure training stability and enhance the network's generative capability. Therefore, we obtain the loss function for k-space as follows:

\[
\begin{aligned}
\mathcal{L}_{k} &= \mathcal{L}_k^{rev}+ \eta \underbrace{\|F_m x^{rec}_p - \tilde{k}\|_1}_{\mathcal{L}_{k}^{reg}},
\end{aligned}
\tag{6} \label{eq13}
\]
where the first term is called the re-visible term and the second term is served as the regularization term. \( \eta \) is a parameter that balances the regularization term and the re-visible term in k-space.

In the image domain, to improve perceptual quality and structural fidelity while leveraging the complementary information between the image domain and k-space, we define the loss function in image domain as follows:
\[
\begin{aligned}
\mathcal{L}_{\text{img}} = & \underbrace{SSIM\left(F^H\frac{{F_m x^{rec}_p + \lambda F_m sg(x^{rec})}}{1 + \lambda} ,  F^H\tilde{k}\right)}_{\mathcal{L}_{\text{img}}^{rev}} \\
& + \eta \underbrace{SSIM(F^HF_m x^{rec}_p, F^H\tilde{k})}_{\mathcal{L}_{\text{img}}^{reg}},
\end{aligned} \tag{7} \label{eq14}
\]
{where $SSIM$ denotes Structural Similarity Index Measure loss \cite{b61}, and widely used in MRI reconstruction \cite{b5,b40}}, $F^H$ denotes inverse fourier transform, \( \lambda \) is a parameter that balances the the regularization term and the re-visible term in image domain. \( \mathcal{L}_{\text{img}}^{rev} \) and \( \mathcal{L}_\text{img}^{reg} \) in image domain correspond to the losses \( \mathcal{L}_k^{rev} \) and \( \mathcal{L}_k^{reg} \) in k-space, further enhancing the model's adaptation to all available data in the image domain.

Finally, we obtain the re-visible dual-domain loss function:
\[
\mathcal{L}_d = \mathcal{L}_{\text{img}} + \beta \mathcal{L}_k, \tag{8} \label{eq15}
\]
where \( \beta \) is used to balance the contributions of the loss in the k-space and image domain. During the training phase, the model has already implicitly learned to utilize all under-sampled k-space data. Therefore, during inference, we directly use all under-sampled k-space data as input to achieve better reconstruction results.

\subsection{Deep Unfolding Network based on CP-PPA}
\label{DUN-CP-PPA}
{Our goal is to construct a deep unfolding network in which each module corresponds to an iterative step of the CP-PPA optimization algorithm described in Section \ref{cpppa}, thereby achieving a more transparent reconstruction process while effectively incorporating image priors.

In MRI reconstruction, based on the settings described in Section \ref{cpppa}, the update rules for the primal and dual variables can be derived as follows:
\[
\left\{
\begin{aligned}
x_{k+1} &= \operatorname{Prox}_f^{\tau_{k+1}} \left( x_k - \tau_{k+1} F_m^H y_k \right), \\
y_{k+1} &= \frac{y_k + \sigma_{k+1} F_m z_{k+1} - \sigma_{k+1} \tilde{k}}{1 + \sigma_{k+1}}.
\end{aligned}
\right.\tag{9} \label{eq9}
\]
Through the iterative updates of \(x\) and \(y\), the imaging physics and image priors are incorporated. The challenge lies in how to design \( \operatorname{Prox}_f^{\tau_{k+1}} \), as deriving it from hand-crafted regularization terms leads to slow convergence and a lack of flexibility. By combining deep unfolding techniques, we unfold this optimization algorithm into a deep unfolding network and name it DUN-CP-PPA.
The pipeline of the proposed DUN-CP-PPA is illustrated in Fig.\ref{fig1}. 
The overall network architecture consists of two key modules: the \textit{Update x Module} and the \textit{Update y Module}. 
The \textit{Update x Module} focuses on learning image-domain priors to refine the reconstruction, 
while the \textit{Update y Module} enforces k-space data consistency by adaptively correcting residual errors. The convergence discussion of DUN-CP-PPA is presented in the Appendix \ref{A1}.

\subsubsection{The Update x Module}
The network strictly follows the iterative form of Eq. \eqref{eq9}. 
This module refines the reconstructed image by incorporating feedback from the dual variable \( y_k \), 
which can be intuitively interpreted as the \( k \)-space residual—representing the inconsistency between the reconstructed image and the acquired \( k \)-space data. 
Specifically, \( y_k \) is first projected into the image domain via the operator \( F_m^{H} \), generating an image-domain residual. 
This residual is then used to adjust \( x_k \), aligning it more closely with the acquired \( k \)-space measurements. 
Finally, the proximal mapping \( \operatorname{Prox}_f^{\tau_{k+1}}(\cdot) \) imposes learned image priors to reduce noise and enhance structural details. 
The resulting update is expressed as  
\(x_{k+1} = \operatorname{Prox}_f^{\tau_{k+1}} (x_k - \tau_{k+1}F_m^{H} y_k).\)

Considering that pre-specified regularization cannot always capture the complex structures of clinically acquired MRI data, 
we adopt a learnable convolutional network, \( \operatorname{ProxNet}_x \), to automatically learn the mapping \( \operatorname{Prox}_f^{\tau_{k+1}}(\cdot) \) 
from training data through our proposed loss backpropagation in an end-to-end manner. 
Previous studies \cite{b9,b10} have demonstrated the effectiveness of the U-Net architecture in extracting multi-scale features 
and its ability to learn denoising priors. 
Therefore, we choose the U-Net architecture as the backbone for \( \operatorname{ProxNet}_x \). 
To capture both global and local feature representations and efficiently learn comprehensive image priors, 
we further utilize a Spatial-Frequency Feature Extraction (SFFE) Block and employ residual connections to enhance training stability. 
This process can be expressed as  
\(x_{k+1} = x_k + \operatorname{ProxNet}_x (x_k - \tau_{k+1}F_m^{H} y_k).\)
The detailed network module design will be introduced in Section \ref{art}.

\subsubsection{The Update y Module}
The network also keeps the iterative structure of Eq. \eqref{eq9}. 
The variable \( z_{k+1} = x_{k+1} + \theta_{k+1}(x_{k+1} - x_k) \) represents the momentum-updated image, which accelerates convergence. 
The operator \( F_m \) maps \( z_{k+1} \) into \( k \)-space to measure the residual inconsistency \( F_m z_{k+1} - \tilde{k} \). 
The adaptive correction is then performed using \( \sigma_{k+1} \), 
adaptively correcting residual errors and finally leading to  
\(y_{k+1} = \frac{y_k + \sigma_{k+1}(F_m z_{k+1} - \tilde{k})}{1 + \sigma_{k+1}}.\)}

\subsubsection{ Architecture of \( \text{ProxNet}_x \)}
\label{art}
The \( \operatorname{ProxNet}_x \) consists of an encoder and a decoder, as shown in Fig. \ref{fig1}. 
The encoder is composed of four SFFE blocks, each containing two feature extraction branches. 
The input feature \( F_{in} \) is fed into both branches simultaneously. 
The first branch uses convolution, instance normalization layers, and Leaky ReLU nonlinearity to learn local features from the image domain. 
In the second branch, inspired by \cite{b44}, the Fourier transform is applied to convert the input \( x \) from the spatial domain to the frequency domain. 
The dot product between the learned frequency-domain features and the learnable global filters is computed to enhance the global features. 
The inverse Fourier transform is then used to convert the learned frequency-domain features back into the spatial domain. 
Finally, both feature representations in the spatial domain are fused using concatenation and convolution operations. 
A residual structure is employed to improve training stability. 
The decoder consists of four Conv \( 3\times3 \) blocks, each containing two convolutional layers with \( 3\times3 \) kernels, 
instance normalization layers, and Leaky ReLU nonlinearity. 
Average pooling and transpose convolutional upsampling are employed to adjust the spatial size of the feature maps. 
Skip connections are utilized to enrich the feature representations.

Through \( k \)-stage optimization, the proposed DUN-CP-PPA is capable of accurately reconstructing the target image. For more details about the network implementation, please refer to Section \ref{ID}. In this way, {by replacing the forward iteration step in the CP-PPA algorithm with a differentiable network module, we naturally construct DUN-CP-PPA based on the derived optimization algorithm. Since DUN-CP-PPA serves as a direct end-to-end mapping from the input (under-sampled data) to the output (fully-sampled data), all step-size parameters and network parameters can be jointly optimized through loss backpropagation of our proposed self-supervised framework in an end-to-end manner. }

By proposing DUN-CP-PPA, the imaging physics and image priors are appropriately incorporated, and the reconstruction process aligns with the iterations of the corresponding algorithm. Therefore, during the training phase, we optimize the algorithm by updating the network parameters using Eq. \eqref{eq15}, ensuring that the model implicitly adapts to all under-sampled k-space data as input while preventing the model from learning the identity mapping. In the inference phase, all under-sampled k-space data can be directly input into the model, leveraging the learned reconstruction process to obtain the final reconstruction results.

\begin{table*}[!ht]
    \centering
    \scriptsize
    \caption{Quantitative evaluation of our method vs. other methods on the fastMRI dataset for T1 and T2 modalities under 4$\times$ and 8$\times$ acceleration with \textbf{equispaced} and \textbf{random} 1D subsampling masks. Best results among self-supervised methods are emphasized in \textbf{bold}. Supervised methods are marked with $^*$.}
    \renewcommand{\arraystretch}{0.96}
    \setlength{\tabcolsep}{7pt}
    \begin{tabular}{l *{9}{c}}
        \toprule
        & \multirow{2}{*}{\textbf{Methods}} & \multicolumn{2}{c}{\textbf{T1 4$\times$ Acceleration}} & \multicolumn{2}{c}{\textbf{T1 8$\times$ Acceleration}} & \multicolumn{2}{c}{\textbf{T2 4$\times$ Acceleration}} & \multicolumn{2}{c}{\textbf{T2 8$\times$ Acceleration}} \\
        \cmidrule(lr){3-4} \cmidrule(lr){5-6} \cmidrule(lr){7-8} \cmidrule(lr){9-10} &
        & \textbf{PSNR} & \textbf{SSIM} & \textbf{PSNR} & \textbf{SSIM} & \textbf{PSNR} & \textbf{SSIM} & \textbf{PSNR} & \textbf{SSIM} \\
        \cmidrule(lr){1-10}
        \multirow{11}{*}{\rotatebox[origin=c]{90}{\textbf{Equispaced}}} &Zero-filling & 27.59$\pm$0.91 & 0.7499$\pm$0.0285 & 24.29$\pm$1.23 & 0.6534$\pm$0.0387 & 26.92$\pm$1.02 & 0.7321$\pm$0.0279 & 24.36$\pm$1.11 & 0.6428$\pm$0.0350 \\
        \multirow{3}{*}{} 
        &SSDU  & 31.79$\pm$0.95 & 0.8925$\pm$0.0160 & 27.55$\pm$0.87 &  0.8218$\pm$0.0194 & 32.42$\pm$1.20 & 0.9082$\pm$0.0099 & 25.81$\pm$0.95 & 0.8025$\pm$0.0207 \\
        &PARCEL  & 32.58$\pm$1.11 & 0.9010$\pm$0.0147 & 29.06$\pm$1.02 &  0.8423$\pm$0.0187 & 33.15$\pm$1.27 & 0.9102$\pm$0.0092 & 28.76$\pm$1.23 & 0.8408$\pm$0.0202 \\
        &SSMRI & 35.69$\pm$1.29 & 0.9238$\pm$0.0121 & 34.76$\pm$0.93 & 0.9163$\pm$0.0119 & 33.75$\pm$1.50 & 0.9122$\pm$0.0119 & 33.12$\pm$1.28 & 0.9168$\pm$0.0125 \\
        &DDSS  & 35.05$\pm$1.17 & 0.9342$\pm$0.0100 & 34.67$\pm$0.91 & 0.9304$\pm$0.0091 & 33.88$\pm$1.20 & 0.9419$\pm$0.0082 & 33.33$\pm$1.38 & 0.9327$\pm$0.0106 \\
        &Noisier2Noise  & 36.02$\pm$1.34 & 0.9384$\pm$0.0104 & 34.87$\pm$0.93 & 0.9244$\pm$0.0105 & 34.06$\pm$1.52 & 0.9350$\pm$0.0122 & 33.49$\pm$1.42 & 0.9337$\pm$0.0122 \\
        & {RSSDU}  & {35.76$\pm$1.30} & {0.9344$\pm$0.0108} & {34.78$\pm$0.94} & {0.9206$\pm$0.0108} & {33.79$\pm$1.51} & {0.9316$\pm$121} & {33.38$\pm$1.36} & {0.9329$\pm$0.0123} \\
        \cmidrule(lr){2-10}
        \multirow{2}{*}{} 
        &U-Net$^*$ & 39.85$\pm$1.38 & 0.9703$\pm$0.0066 & 37.61$\pm$1.47 &  0.9579$\pm$0.0095 & 35.85$\pm$1.55 & 0.9543$\pm$0.0102 & 33.90$\pm$1.68 & 0.9396$\pm$0.0137 \\
        &Varnet$^*$ & 41.75$\pm$1.44 & 0.9813$\pm$0.0045 & 39.91$\pm$1.51 & 0.9740$\pm$0.0064 & 38.49$\pm$1.60 & 0.9751$\pm$0.0067 & 36.63$\pm$1.66 & 0.9658$\pm$0.0088 \\
        \cmidrule(lr){2-10}
        \multirow{1}{*}{} 
        &Ours  & \textbf{39.16$\pm$1.08} & \textbf{0.9674$\pm$0.0058} & \textbf{37.35$\pm$1.39} & \textbf{0.9539$\pm$0.0089} & \textbf{36.97$\pm$1.32} & \textbf{0.9642$\pm$0.0070} & \textbf{35.17$\pm$1.52} & \textbf{0.9502$\pm$0.0104} \\
        \cmidrule(lr){1-10}
        \multirow{11}{*}{\rotatebox[origin=c]{90}{\textbf{Random}}} &Zero-filling & 27.41$\pm$0.92 & 0.7518$\pm$0.0281 & 24.19$\pm$1.24 & 0.6739$\pm$0.0371 & 26.77$\pm$1.02 & 0.7332$\pm$0.0282 & 24.26$\pm$1.13 & 0.6669$\pm$0.0345  \\
        \multirow{3}{*}{} 
        &SSDU  & 34.40$\pm$1.36 & 0.9178$\pm$0.0134 & 27.66$\pm$1.32 & 0.8275$\pm$0.0197 & 32.18$\pm$1.20 & 0.9116$\pm$0.0119 & 25.38$\pm$1.03 & 0.7925$\pm$0.0212 \\
        &PARCEL  & 37.06$\pm$1.31 & 0.9357$\pm$0.0125 & 29.73$\pm$1.41 &  0.8456$\pm$0.0236 & 35.21$\pm$1.35 & 0.9321$\pm$0.0140 & 26.77$\pm$1.34 & 0.8073$\pm$0.0228 \\
        &SSMRI & 38.49$\pm$1.58 & 0.9429$\pm$0.0082 & 30.14$\pm$2.05 & 0.9002$\pm$0.0143 & 37.67$\pm$1.56 & 0.9312$\pm$0.0104 & 31.68$\pm$1.48 & 0.9040$\pm$0.0124 \\
        &DDSS  & 40.48$\pm$1.29 & 0.9703$\pm$0.0055 & 33.09$\pm$1.36 & 0.9223$\pm$0.0117 & 36.95$\pm$1.41 & 0.9552$\pm$0.0083 & 31.33$\pm$1.44 & 0.9156$\pm$0.0130 \\
        &Noisier2Noise  & 39.55$\pm$1.45 & 0.9603$\pm$0.0083 & 30.39$\pm$1.46 & 0.9059$\pm$0.0128 & 38.58$\pm$1.55 & 0.9640$\pm$0.0087 & 31.77$\pm$1.50 & 0.9184$\pm$0.0145 \\
        &{RSSDU}  & {38.79$\pm$1.49} & {0.9517$\pm$0.0082} & {30.27$\pm$1.58} & {0.9028$\pm$0.0132} & {37.86$\pm$1.58} & {0.9591$\pm$0.0096} & {31.58$\pm$1.48} & {0.9032$\pm$0.0138} \\

        \cmidrule(lr){2-10}
        \multirow{2}{*}{} 
        &U-Net$^*$ & 40.65$\pm$1.36 & 0.9723$\pm$0.0064 & 35.45$\pm$1.61 &  0.9469$\pm$0.0120 & 36.63$\pm$1.47 & 0.9570$\pm$0.0097 & 32.30$\pm$1.59 & 0.9285$\pm$0.0147 \\
        &Varnet$^*$ & 44.40$\pm$1.49 & 0.9873$\pm$0.0032 & 38.12$\pm$1.58 & 0.9675$\pm$0.0077 & 41.60$\pm$1.76 & 0.9842$\pm$0.0049 & 35.29$\pm$1.70 & 0.9588$\pm$0.0105 \\
        \cmidrule(lr){2-10}
        \multirow{1}{*}{} 
        &Ours  & \textbf{42.56$\pm$1.39} & \textbf{0.9802$\pm$0.0040} & \textbf{34.39$\pm$1.52} & \textbf{0.9338$\pm$0.0118} & \textbf{40.03$\pm$1.58} & \textbf{0.9764$\pm$0.0057} & \textbf{32.85$\pm$1.53} & \textbf{0.9324$\pm$0.0130} \\
        
        \bottomrule
    \end{tabular}
    \label{tb1}
\end{table*}

\begin{figure*}[!t]
\centering 
\includegraphics[width=0.98\textwidth]{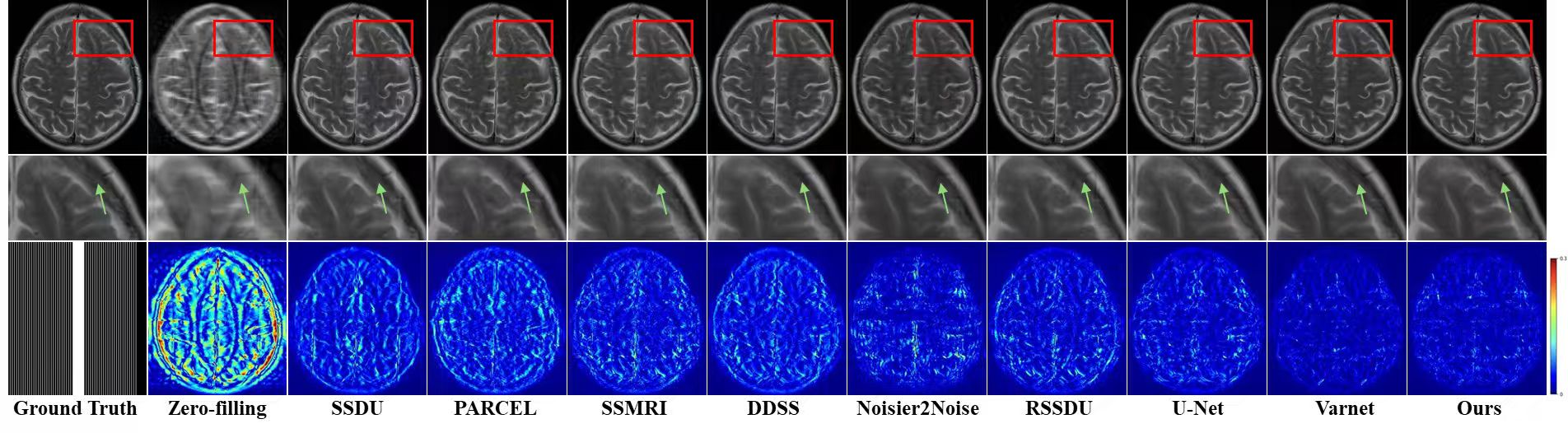} 
\caption{Visual comparison of methods for 4$\times$ acceleration using a 1D equispaced subsampling mask on fastMRI dataset. First row: Reconstructed images; second row: Zoomed region; third row: Mask and error maps.}
\label{fig3}
\end{figure*}

\section{Experiments}
\subsection{Datasets}
{We evaluate our method on both single-coil and multi-coil datasets. 

In single-coil setting, We use the single-coil fastMRI brain dataset and IXI dataset. For \textbf{single-coil fastMRI brain dataset \footnote{https://fastMRI.med.nyu.edu/.}}, we adopt the configurations outlined in \cite{b40}. A total of 340 paired {Longitudinal relaxation time–weighted (T1-weighted)} and {Transverse relaxation time–weighted (T2-weighted)} axial brain MRI scans are selected and divided into three subsets: 170 volumes (comprising 2720 slices) for training, 68 volumes (1088 slices) for validation, and 102 volumes (1632 slices) for testing. Each T1-weighted and T2-weighted image has an in-plane dimension of \(1 \times320 \times 320\). For \textbf{IXI Dataset\footnote{http://brain-development.org/ixi-dataset/.}}, we select 570 pairs of Proton density–weighted (PD-weighted) and T2-weighted axial brain MRIs for our study which are divided into 285 volumes for training, 115 volumes for validation, and 170 volumes for testing. Each PD-weighted and T2-weighted image has an in-plane resolution of \(1 \times 256 \times 256\). Following the approach described in \cite{b42}, we utilize the central 100 slices from each volume in our experiments.

In multi-coil setting, we use the multi-coil fastMRI brain dataset and multi-coil knee dataset from the Center for Advanced Imaging Innovation and Research. For \textbf{Multi-coil fastMRI brain dataset \footnote{{https://fastMRI.med.nyu.edu/.}}}, a total of 90 T2-weighted axial 20-coil brain MRI volumes are selected and divided into three subsets: 72 volumes (comprising 1152 slices) for training, 9 volumes (144 slices) for validation, and 9 volumes (144 slices) for testing. The slices are extracted from each subject with a dimension of \(20 \times 320 \times 320\). For \textbf{Multi-coil knee dataset\footnote{{https://cai2r.net/resources/100-
knee-mri-cases/.}}}, a total of 20 PD-weighted 15-coil knee MRI volumes are selected and divided into three subsets: 15 volumes for training, 2 volumes for validation, and 3 volumes for testing. The slices are extracted from each subject with a dimension of \(15 \times 320 \times 320\).} We utilize the central 30 slices from each volume in our experiments.

\subsection{Compared Methods}
\begin{table*}[!ht]
    \centering
    \scriptsize
    \caption{Quantitative evaluation of our method vs. other methods on the IXI dataset for PD and T2 modalities under 4$\times$ and 8$\times$ acceleration with \textbf{equispaced} and \textbf{random} 1D subsampling masks. Best results among self-supervised methods are emphasized in \textbf{bold}. Supervised methods are marked with $^*$.}
    \renewcommand{\arraystretch}{0.96}
    \setlength{\tabcolsep}{7pt}
    \begin{tabular}{l *{9}{c}}
        \toprule
        & \multirow{2}{*}{\textbf{Methods}} & \multicolumn{2}{c}{\textbf{PD 4$\times$ Acceleration}} & \multicolumn{2}{c}{\textbf{PD 8$\times$ Acceleration}} & \multicolumn{2}{c}{\textbf{T2 4$\times$ Acceleration}} & \multicolumn{2}{c}{\textbf{T2 8$\times$ Acceleration}} \\
        \cmidrule(lr){3-4} \cmidrule(lr){5-6} \cmidrule(lr){7-8} \cmidrule(lr){9-10} &
        & \textbf{PSNR} & \textbf{SSIM} & \textbf{PSNR} & \textbf{SSIM} & \textbf{PSNR} & \textbf{SSIM} & \textbf{PSNR} & \textbf{SSIM} \\
        \cmidrule(lr){1-10}
        \multirow{11}{*}{\rotatebox[origin=c]{90}{\textbf{Equispaced}}} &Zero-filling & 26.96$\pm$2.45 & 0.6337$\pm$0.0567 & 24.26$\pm$2.48 & 0.5442$\pm$0.0700 & 26.56$\pm$2.04 & 0.6337$\pm$0.0449 & 24.08$\pm$2.01 & 0.5517$\pm$0.0535 \\
        \multirow{3}{*}{} 
        &SSDU  & 35.69$\pm$1.79 & 0.9397$\pm$0.0072 & 28.47$\pm$1.78 & 0.8120$\pm$0.0181 & 34.30$\pm$1.32 & 0.9384$\pm$0.0094 & 28.27$\pm$1.84 & 0.8422$\pm$0.0165 \\
        &PARCEL  & 37.05$\pm$2.26 & 0.9483$\pm$0.0074 & 29.75$\pm$2.10 &  0.8378$\pm$0.0202 & 37.35$\pm$1.99 & 0.9516$\pm$0.0099 & 29.86$\pm$1.93 & 0.8531$\pm$0.0183 \\
        &SSMRI & 36.96$\pm$2.40 & 0.9457$\pm$0.0087 & 31.78$\pm$2.37 & 0.8657$\pm$0.0206 & 37.07$\pm$2.06 & 0.9461$\pm$0.0087 & 31.86$\pm$2.04 & 0.8607$\pm$0.0223 \\
        &DDSS  & 39.90$\pm$2.48 & 0.9762$\pm$0.0069 & 31.08$\pm$2.39 & 0.8896$\pm$0.0214 & 38.91$\pm$2.21 & 0.9722$\pm$0.0072 & 30.44$\pm$2.08 & 0.8930$\pm$0.0183 \\
        &Noisier2Noise  & 38.78$\pm$2.40 & 0.9571$\pm$0.0130 & 31.86$\pm$2.37 & 0.9010$\pm$0.0222 & 39.12$\pm$2.14 & 0.9641$\pm$0.0093 & 32.21$\pm$2.03 & 0.9138$\pm$0.0165 \\
        &{RSSDU}  & {37.44$\pm$2.39} & {0.9516$\pm$0.0088} & {31.67$\pm$2.38} & {0.8934$\pm$0.0215} & {38.26$\pm$2.12} & {0.9569$\pm$0.0086} & {31.98$\pm$2.01} & {0.9042$\pm$0.0192} \\
        \cmidrule(lr){2-10}
        \multirow{2}{*}{} 
        &U-Net$^*$ & 38.83$\pm$2.54 & 0.9715$\pm$0.0087 & 32.74$\pm$2.42 & 0.9210$\pm$0.0189 & 38.44$\pm$2.30 & 0.9694$\pm$0.0086 & 31.69$\pm$2.19 & 0.9127$\pm$0.0184 \\
        &Varnet$^*$ & 42.18$\pm$2.53 & 0.9866$\pm$0.0041 & 35.04$\pm$2.35 & 0.9544$\pm$0.0107 & 41.61$\pm$2.30 & 0.9854$\pm$0.0046 & 34.31$\pm$2.09 & 0.9509$\pm$0.0110 \\
        \cmidrule(lr){2-10}
        \multirow{1}{*}{} 
        &Ours  & \textbf{41.64$\pm$2.46} & \textbf{0.9834$\pm$0.0049} & \textbf{33.19$\pm$2.36} & \textbf{0.9254$\pm$0.0164} & \textbf{41.36$\pm$2.20} & \textbf{0.9819$\pm$0.0053} & \textbf{33.22$\pm$2.04} & \textbf{0.9304$\pm$0.0142} \\
        \cmidrule(lr){1-10}
        \multirow{11}{*}{\rotatebox[origin=c]{90}{\textbf{Random}}} &Zero-filling & 26.80$\pm$2.45 & 0.6345$\pm$0.0595 & 23.99$\pm$2.48 & 0.5409$\pm$0.0710 & 26.40$\pm$2.04 & 0.6389$\pm$0.0470 & 23.83$\pm$2.01 & 0.5493$\pm$0.0539 \\
        \multirow{3}{*}{} 
        &SSDU  & 34.99$\pm$1.75 & 0.9328$\pm$0.0093 & 28.14$\pm$1.91 & 0.8346$\pm$0.0172 & 32.78$\pm$1.60 & 0.9076$\pm$0.0107 & 27.58$\pm$1.86 & 0.8327$\pm$0.0168 \\
        &PARCEL  & 35.05$\pm$2.30 & 0.9344$\pm$0.0093 & 28.77$\pm$2.14 &  0.8396$\pm$0.0261 & 34.26$\pm$1.93 & 0.9167$\pm$0.0160 & 28.91$\pm$1.97 & 0.8425$\pm$0.0217 \\
        &SSMRI & 35.98$\pm$2.37 & 0.9389$\pm$0.0092 & 30.36$\pm$2.38 & 0.8385$\pm$0.0201 & 34.46$\pm$2.02 & 0.9114$\pm$0.0120 & 30.66$\pm$2.02 & 0.8382$\pm$0.0227 \\
        &DDSS  & 35.43$\pm$2.41 & 0.9449$\pm$0.0135 & 29.08$\pm$2.39 & 0.8378$\pm$0.0290 & 35.38$\pm$2.10 & 0.9489$\pm$0.0110 & 28.70$\pm$2.05 & 0.8582$\pm$0.0225 \\
        &Noisier2Noise  & 36.89$\pm$2.39 & 0.9537$\pm$0.0128 & 30.52$\pm$2.38 & 0.8774$\pm$0.0265 & 35.06$\pm$2.05 & 0.9339$\pm$0.0146 & 30.78$\pm$2.01 & 0.8942$\pm$0.0193 \\
        &{RSSDU}  & {36.21$\pm$2.38} & {0.9477$\pm$0.0117} & {30.48$\pm$2.38} & {0.8724$\pm$0.0245} & {34.76$\pm$2.03} & {0.9289$\pm$0.0128} & {30.68$\pm$2.01} & {0.8862$\pm$0.0207} \\
        \cmidrule(lr){2-10}
        \multirow{2}{*}{} 
        &U-Net$^*$ & 35.79$\pm$2.45 & 0.9482$\pm$0.0143 & 31.90$\pm$2.39 & 0.9080$\pm$0.0215 & 34.67$\pm$2.16 & 0.9407$\pm$0.0136 & 30.57$\pm$2.16 & 0.8945$\pm$0.0210 \\
        &Varnet$^*$ & 39.82$\pm$2.39 & 0.9804$\pm$0.0055 & 34.13$\pm$2.33 & 0.9465$\pm$0.0120 & 38.83$\pm$2.14 & 0.9764$\pm$0.0063 & 33.51$\pm$2.05 & 0.9433$\pm$0.0121 \\
        \cmidrule(lr){2-10}
        \multirow{1}{*}{} 
        &Ours  & \textbf{38.81$\pm$2.39} & \textbf{0.9730$\pm$0.0072} & \textbf{31.37$\pm$2.33} & \textbf{0.9025$\pm$0.0203} & \textbf{38.21$\pm$2.06} & \textbf{0.9706$\pm$0.0072} & \textbf{32.17$\pm$1.97} & \textbf{0.9168$\pm$0.0160} \\
        
        \bottomrule
    \end{tabular}
    \label{tb2}
\end{table*}

\begin{figure*}[!t]
\centering 
\includegraphics[width=0.98\textwidth]{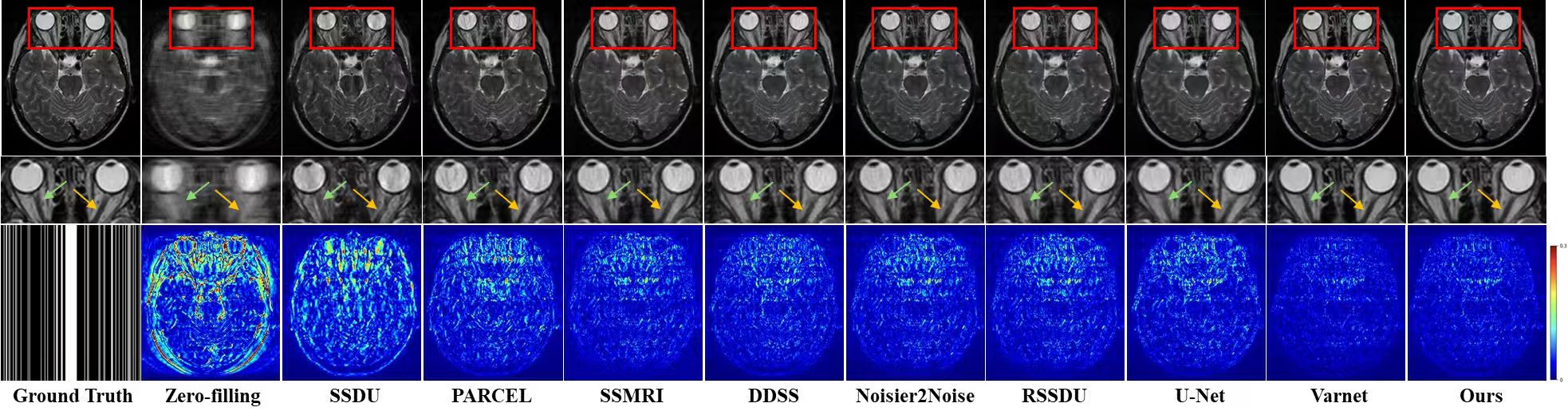} 
\caption{Visual comparison of methods for 4$\times$ acceleration using a 1D random subsampling mask on IXI dataset. First row: Reconstructed images; second row: Zoomed region; third row: Mask and error maps.}
\label{fig4}
\end{figure*}
In single-coil setting, we evaluate our proposed method by comparing it with 
{eight} methods: two supervised methods (U-Net, E2E-Varnet\cite{b5}) and six self-supervised methods (SSDU\cite{b32}, PARCEL\cite{b45}, SSMRI \cite{b15}, DDSS \cite{b18}, Noisier2Noise\cite{b46}, RSSDU\cite{b53}).  
{In multi-coil setting, we evaluate our proposed method by comparing it with seven methods: two supervised methods (U-Net, E2E-Varnet\cite{b5}) and five self-supervised methods (SSMRI \cite{b15}, DDSS \cite{b18}, Noisier2Noise\cite{b46}, RSSDU\cite{b53}, AeSPa\cite{b54})}. U-Net consists of four encoder and four decoder and incorporates a data consistency block to enhance reconstruction performance. E2E-Varnet applies the variational technique, combining it with U-Net to unfold the process into an end-to-end learning framework. SSDU {utilizes} loss defined on all the scanned data points and adopts a parallel training framework for self-supervised MRI reconstruction. PARCEL combines the advantages of contrastive representation learning with model-based deep learning MRI reconstruction models to form an efficient reconstruction strategy. SSMRI uses VarNet as its backbone and apply the diagonal weight mask to adjust loss function for better training. DDSS introduces a triple-branch, dual-domain self-supervised reconstruction approach. Noisier2Noise proposes a loss weighting that compensates for the sampling and partitioning
densities. {RSSDU resamples k-space data twice with different acceleration factors and trains a network to map one subset to the other. AeSPa is an attention-guided self-supervised parallel imaging method with cross-channel constraint to achieve robust sensitivity map estimation.}

\subsection{Implementation Details}
\label{ID}
In our experiments, we follow the fastMRI challenge paradigm, generating under-sampled MRIs by masking fully-sampled k-space data using Cartesian patterns. Two sampling strategies are employed: random and equispaced, at sampling ratios of 25\% (4$\times$ acceleration) and 12.5\% (8$\times$ acceleration). To exploit the high-energy content in low-frequency k-space, 32\% of the samples are allocated to these frequencies, with the rest distributed either randomly or equispaced. For hyperparameter selection, we set the model parameters as stage$=8$, $\lambda=10$, $\eta=1$, $\beta=10$ for our experiments. All models are implemented in PyTorch and trained on a system with four NVIDIA GeForce GTX 3090 GPUs. The batch size is set to \(2\), and parameters are optimized using the Adam optimizer with a learning rate of \(1 \times 10^{-4}\). A validation set is used to prevent overfitting and select optimal parameters, which are then applied to the test set for final evaluation. During training, the under-sampled data partitioning rate is randomly generated between $[0.2, 0.8]$ which was empirically determined to provide better results. The experimental results are all based on reconstructed 3D data.

\subsection{Evaluation of Proposed Method}
\begin{table*}[!ht]
    \centering
    \scriptsize
    \caption{{Quantitative evaluation of our method vs. other methods on the multi-coil brain and knee datasets under 4$\times$ and 8$\times$ acceleration with \textbf{equispaced} and \textbf{random} 1D subsampling masks. Best results among self-supervised methods are emphasized in \textbf{bold}. Supervised methods are marked with $^*$.}}
    \renewcommand{\arraystretch}{0.96}
    \setlength{\tabcolsep}{7pt}
    {\begin{tabular}{l *{9}{c}}
        \toprule
        & \multirow{2}{*}{\textbf{Methods}} & \multicolumn{2}{c}{\textbf{Knee PD 4$\times$ Acceleration}} & \multicolumn{2}{c}{\textbf{Knee PD 8$\times$ Acceleration}} & \multicolumn{2}{c}{\textbf{Brain 4$\times$ T2 Acceleration}} & \multicolumn{2}{c}{\textbf{Brain 8$\times$ T2 Acceleration}} \\
        \cmidrule(lr){3-4} \cmidrule(lr){5-6} \cmidrule(lr){7-8} \cmidrule(lr){9-10} &
        & \textbf{PSNR} & \textbf{SSIM} & \textbf{PSNR} & \textbf{SSIM} & \textbf{PSNR} & \textbf{SSIM} & \textbf{PSNR} & \textbf{SSIM} \\
        \cmidrule(lr){1-10}
        \multirow{10}{*}{\rotatebox[origin=c]{90}{\textbf{Equispaced}}} &Zero-filling & 28.79$\pm$1.61 & 0.7585$\pm$0.0399 & 24.93$\pm$1.30 & 0.6909$\pm$0.0380 & 27.43$\pm$1.29 & 0.7720$\pm$0.0276 & 24.46$\pm$1.43 & 0.7190$\pm$0.0369 \\
        &SSMRI  & 32.27$\pm$1.64 & 0.8379$\pm$0.0288 & 30.27$\pm$1.54 & 0.8103$\pm$0.0311 & 31.00$\pm$1.78 & 0.8686$\pm$0.0379 & 30.36$\pm$1.61 & 0.8491$\pm$0.0328 \\
        \multirow{3}{*}{} 
        &DDSS  & 32.66$\pm$1.62 & 0.8549$\pm$0.0276 & 30.84$\pm$1.57 & 0.8212$\pm$0.0336 & 32.96$\pm$1.64 & 0.8946$\pm$0.0353 & 31.02$\pm$1.73 & 0.8688$\pm$0.0352 \\
        &Noisier2Noise  & 32.83$\pm$1.75 & 0.8468$\pm$0.0307 & 30.96$\pm$1.78 &  0.8189$\pm$0.0367 & 32.39$\pm$1.52 & 0.8876$\pm$0.0371 & 30.93$\pm$1.63 & 0.8647$\pm$0.0333 \\
        &RSSDU  & 32.54$\pm$1.67 & 0.8427$\pm$0.0302 & 30.46$\pm$1.69 & 0.8148$\pm$0.0379 & 31.29$\pm$1.59 & 0.8796$\pm$0.0326 & 30.79$\pm$1.70 & 0.8597$\pm$0.0356 \\
        &AeSPa  & 33.01$\pm$1.78 & 0.8577$\pm$0.0284 & 31.18$\pm$1.63 & 0.8201$\pm$0.0355 & 33.21$\pm$1.88 & 0.8976$\pm$0.0338 & 30.74$\pm$1.68 & 0.8573$\pm$0.0354 \\
        \cmidrule(lr){2-10}
        \multirow{2}{*}{} 
        &U-Net$^*$ & 32.16$\pm$1.61 & 0.8526$\pm$0.0242 & 30.64$\pm$2.42 & 0.8209$\pm$0.0189 & 32.49$\pm$1.72 & 0.8830$\pm$0.0363 & 30.69$\pm$2.03 & 0.8585$\pm$0.0381 \\
        &Varnet$^*$ & 34.83$\pm$1.91 & 0.8803$\pm$0.0235 & 32.64$\pm$1.53 & 0.8504$\pm$0.0301 & 35.29$\pm$1.95 & 0.9159$\pm$0.0342 & 32.85$\pm$1.93 & 0.8981$\pm$0.0405 \\
        \cmidrule(lr){2-10}
        \multirow{1}{*}{} 
        &Ours  & \textbf{33.71$\pm$1.87} & \textbf{0.8637$\pm$0.0302} & \textbf{31.80$\pm$1.79} & \textbf{0.8329$\pm$0.0360} & \textbf{34.07$\pm$1.90} & \textbf{0.9024$\pm$0.0415} & \textbf{31.96$\pm$1.75} & \textbf{0.8819$\pm$0.0369} \\
        \cmidrule(lr){1-10}
        \multirow{10}{*}{\rotatebox[origin=c]{90}{\textbf{Random}}} &Zero-filling & 28.72$\pm$1.61 & 0.7863$\pm$0.0345 & 24.88$\pm$1.34 & 0.6972$\pm$0.0379 & 27.37$\pm$1.30 & 0.7997$\pm$0.0261 & 24.32$\pm$1.45 & 0.7252$\pm$0.0382 \\
        &SSMRI  & 34.72$\pm$1.82 & 0.8867$\pm$0.0227 & 29.11$\pm$1.73 & 0.7816$\pm$0.0386 & 34.77$\pm$1.87 & 0.9118$\pm$0.0290 & 28.80$\pm$1.70 & 0.8317$\pm$0.0375 \\
        \multirow{3}{*}{} 
        &DDSS  & 35.50$\pm$1.96 & 0.9101$\pm$0.0291 & 29.59$\pm$1.73 & 0.8076$\pm$0.0358 & 35.39$\pm$2.09 & 0.9170$\pm$0.0333 & 29.06$\pm$1.80 & 0.8431$\pm$0.0376 \\
        &Noisier2Noise  & 35.76$\pm$1.89 & 0.9062$\pm$0.0261 & 29.85$\pm$1.73 &  0.8056$\pm$0.0341 & 35.49$\pm$2.30 & 0.9118$\pm$0.0416 & 29.37$\pm$1.86 & 0.8398$\pm$0.0424 \\
        &RSSDU  & 35.15$\pm$1.80 & 0.9041$\pm$0.0203 & 29.60$\pm$1.73 & 0.8038$\pm$0.0345 & 35.05$\pm$1.95 & 0.9151$\pm$0.0307 & 29.27$\pm$1.74 & 0.8356$\pm$0.0386 \\
        &AeSPa  & 35.84$\pm$1.81 & 0.9107$\pm$0.0229 & 29.38$\pm$1.78 & 0.8018$\pm$0.0214 & 35.31$\pm$2.23 & 0.9162$\pm$0.0323 & 29.14$\pm$1.81 & 0.8322$\pm$0.0392 \\
        \cmidrule(lr){2-10}
        \multirow{2}{*}{} 
        &U-Net$^*$ & 34.62$\pm$1.59 & 0.8986$\pm$0.0192 & 29.57$\pm$1.31 & 0.8027$\pm$0.0358 & 34.32$\pm$2.12 & 0.9107$\pm$0.0302 & 29.18$\pm$2.06 & 0.8409$\pm$0.0479 \\
        &Varnet$^*$ & 37.36$\pm$1.98 & 0.9286$\pm$0.0183 & 31.85$\pm$1.80 & 0.8310$\pm$0.0297 & 37.33$\pm$2.30 & 0.9344$\pm$0.0311 & 31.46$\pm$2.04 & 0.8779$\pm$0.0462 \\
        \cmidrule(lr){2-10}
        \multirow{1}{*}{} 
        &Ours  & \textbf{36.57$\pm$1.94} & \textbf{0.9181$\pm$0.0200} & \textbf{30.53$\pm$1.73} & \textbf{0.8163$\pm$0.0348} & \textbf{36.13$\pm$2.33} & \textbf{0.9230$\pm$0.0410} & \textbf{30.01$\pm$1.84} & \textbf{0.8560$\pm$0.0412} \\
        
        \bottomrule
    \end{tabular}}
    \label{tb100}
\end{table*}

\begin{figure*}[!t]
\centering 
\includegraphics[width=0.98\textwidth]{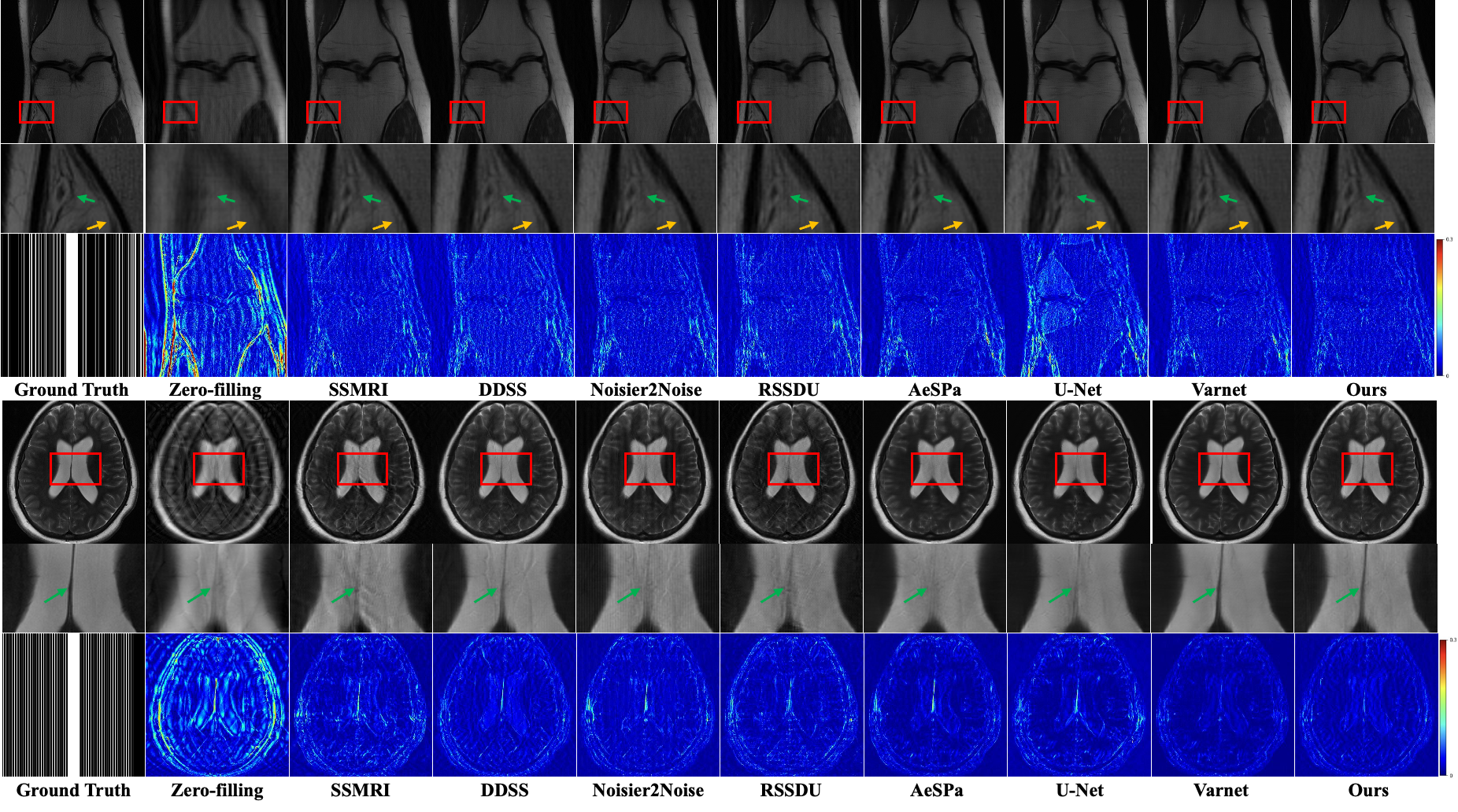} 
\caption{{Visual comparison of different methods for 4$\times$ acceleration using a 1D random subsampling mask on the multi-coil knee dataset and 4$\times$ acceleration using a 1D equispaced subsampling mask on the multi-coil brain dataset. For each dataset, results are displayed as follows: the first row shows the reconstructed images, the second row presents the zoomed-in regions, and the third row illustrates the sampling masks and corresponding error maps.}}
\label{fig100}
\end{figure*}

\begin{figure*}[!t]
\centering 
\includegraphics[width=\textwidth]{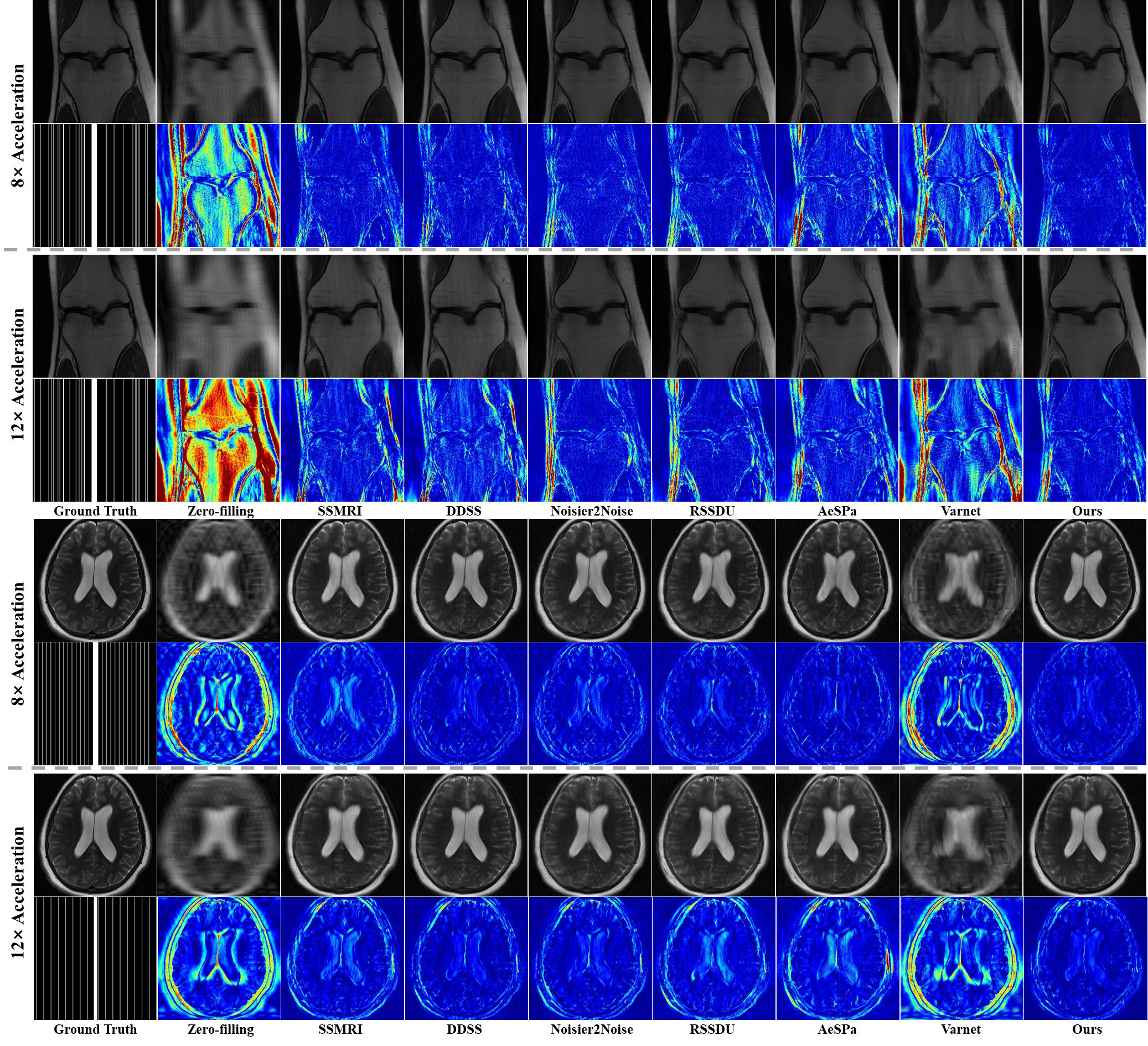} 
\caption{{Visual comparison of methods for generalization performance on the multi-coil brain and knee datasets in scenarios of pattern shift. The model is trained at 4$\times$ acceleration and directly tested at 8$\times$ and 12$\times$ acceleration.}}
\label{fig8}
\end{figure*}
\subsubsection{Comparison with State-of-the-Arts}
In Table \ref{tb1}, we present the quantitative comparisons in fastMRI dataset for T1- and T2-weighted image reconstructions under equispaced and random subsampling masks with 4$\times$ and 8$\times$ accelerations. The results demonstrate that our method achieves significant improvements in PSNR and SSIM compared to other self-supervised methods. Compared to supervised methods, our approach—powered by the advanced DUN-CP-PPA and a re-visible dual-domain self-supervised approach—outperforms the U-Net supervised model, though there remains a performance gap when compared to the Varnet supervised model. Additionally, in Fig. \ref{fig3}, we provide a visual comparison of different approaches for T2-weighted images on the fastMRI dataset under equispaced subsampling with 4$\times$ acceleration. The second row highlights the zoomed-in region of interest (ROI), while the third row displays the corresponding error maps. From the close-up ROI and error maps, it is evident that our method achieves superior visual quality among all self-supervised methods, particularly for fine-grained structures, as indicated by the yellow arrow and blue arrow.

We further validate the proposed method on the public IXI dataset. In Fig. \ref{fig4} and Table \ref{tb2}, we report performance evaluations under 1D equispaced and random masks with 4$\times$ and 8$\times$ accelerations for both PD-weighted and T2-weighted images. Consistently, our method outperforms competing approaches qualitatively and quantitatively, aligning with the observed performance improvements on the fastMRI dataset. This indicates the strong generalizability of the proposed framework.

{In addition, we conduct experiments on the multi-coil brain and knee datasets. To adapt to the multi-coil setting, we employed the same strategy as E2E-VarNet. Specifically, the difference from the single-coil case lies in replacing $F_m = MF$ with $\mathcal{E} = M F \mathcal{S}$, where the newly introduced operator \(\mathcal{S} : x \mapsto [S_c x]_c\) represents multiplying the image by each coil sensitivity $S_c$ and stacking the results across channels. Correspondingly, in the network design, we use an additional lightweight network to predict coil sensitivities, following the approach used in E2E-VarNet. As shown in Figure \ref{fig100} and Table \ref{tb100}, we report performance evaluations under 1D equispaced and random masks with 4$\times$ and 8$\times$ accelerations for both PD-weighted knees and T2-weighted brains. Our method again achieves the best results among all compared methods in both qualitative and quantitative evaluations. These findings are consistent with those on the single-coil dataset, confirming the general effectiveness and robustness of our approach.}

\subsubsection{Generalization Performance in scenarios of Pattern Shift}
Generalization ability is essential for the practical application of deep learning-based methods. To evaluate this, we design experiments to test model performance under varying sampling patterns during training and testing {in multi-coil knee and brain datasets.} Specifically, the model is trained with 4$\times$ acceleration and tested with 8$\times$ and 12$\times$ acceleration. As shown in Table \ref{tb8} and Fig. \ref{fig8}, supervised learning methods show significant performance degradation, demonstrating their poor generalization capability to pattern shift. Self-supervised learning methods all outperform supervised methods, indicating better generalization performance to pattern shift. Among all self-supervised learning approaches, our method achieved the best SSIM and PSNR metrics, as well as superior image detail restoration, in scenarios involving pattern shift.

\begin{table}[!ht]
    \centering
    \scriptsize
    \renewcommand{\arraystretch}{0.96}
    \setlength\tabcolsep{8pt}
    \caption{{Evaluation of Generalization Performance under Pattern Shift Scenarios on the Multi-Coil Brain and knee Datasets.}}
    \label{tb8}
    {\begin{tabular}{cccc}
        \toprule
        & \textbf{Method} & \textbf{8$\times$ Acceleration} & \textbf{12$\times$ Acceleration} \\
        \midrule
        \multirow{14}{*}{\rotatebox[origin=c]{90}{\textbf{Knee}}} 
        & \multirow{2}{*}{SSMRI} & 30.21 $\pm$ 1.78 & 26.97 $\pm$ 1.72 \\
        & & 0.8487 $\pm$ 0.0358 & 0.7966 $\pm$ 0.0422 \\
        & \multirow{2}{*}{DDSS} & 30.89 $\pm$ 1.84 & 27.45 $\pm$ 1.66 \\
        & & 0.8742 $\pm$ 0.0380 & 0.8186 $\pm$ 0.0430 \\
        & \multirow{2}{*}{Noisier2Noise} & 30.75 $\pm$ 1.75 & 27.34 $\pm$ 1.65 \\
        & & 0.8690 $\pm$ 0.0355 & 0.8147 $\pm$ 0.0413 \\
        & \multirow{2}{*}{RSSDU} & 30.49 $\pm$ 1.73 & 27.15 $\pm$ 1.66 \\
        & & 0.8623 $\pm$ 0.0350 & 0.8114 $\pm$ 0.0154 \\
        & \multirow{2}{*}{AeSPa} & 30.74 $\pm$ 1.68 & 26.88 $\pm$ 1.49 \\
        & & 0.8573 $\pm$ 0.0354 & 0.7923 $\pm$ 0.0433 \\
        & \multirow{2}{*}{Varnet$^*$} & 24.57 $\pm$ 1.51 & 23.08 $\pm$ 1.50 \\
        & & 0.7056 $\pm$ 0.0433 & 0.6421 $\pm$ 0.0489 \\
        & \multirow{2}{*}{Ours} & \textbf{32.04 $\pm$ 1.84} & \textbf{28.43 $\pm$ 1.69} \\
        & & \textbf{0.8801 $\pm$ 0.0388} & \textbf{0.8305 $\pm$ 0.0462} \\
        \midrule
        \multirow{14}{*}{\rotatebox[origin=c]{90}{\textbf{Brain}}} 
        & \multirow{2}{*}{SSMRI} & 30.72 $\pm$ 1.77 & 27.96 $\pm$ 1.32 \\
        & & 0.8361 $\pm$ 0.0317 & 0.7842 $\pm$ 0.0304 \\
        & \multirow{2}{*}{DDSS} & 31.38 $\pm$ 1.71 & 28.26 $\pm$ 1.21 \\
        & & 0.8481 $\pm$ 0.0301 & 0.7966 $\pm$ 0.0286 \\
        & \multirow{2}{*}{Noisier2Noise} & 31.54 $\pm$ 1.73 & 28.56 $\pm$ 1.44 \\
        & & 0.8479 $\pm$ 0.0327 & 0.8012 $\pm$ 0.0326 \\
        & \multirow{2}{*}{RSSDU} & 31.04 $\pm$ 1.72 & 28.13 $\pm$ 1.36 \\
        & & 0.8405 $\pm$ 0.0327  & 0.7897 $\pm$ 0.0322 \\
        & \multirow{2}{*}{AeSPa} & 29.38 $\pm$ 1.78 & 27.66 $\pm$ 1.39 \\
        & & 0.8018 $\pm$ 0.0214 & 0.7673 $\pm$ 0.0331 \\
        & \multirow{2}{*}{Varnet$^*$} & 27.11 $\pm$ 1.36 & 24.08 $\pm$ 1.27 \\
        & & 0.7536 $\pm$ 0.0338 & 0.6572 $\pm$ 0.0370 \\
        & \multirow{2}{*}{Ours} & \textbf{32.39 $\pm$ 1.89} & \textbf{29.28 $\pm$ 1.45} \\
        & & \textbf{0.8598 $\pm$ 0.0315} & \textbf{0.8122 $\pm$ 0.0324} \\        
        \bottomrule
    \end{tabular}}
\end{table}

\subsubsection{Generalization Performance under noisy conditions}
\begin{figure}[ht]
\centering 
\includegraphics[width=0.48\textwidth]{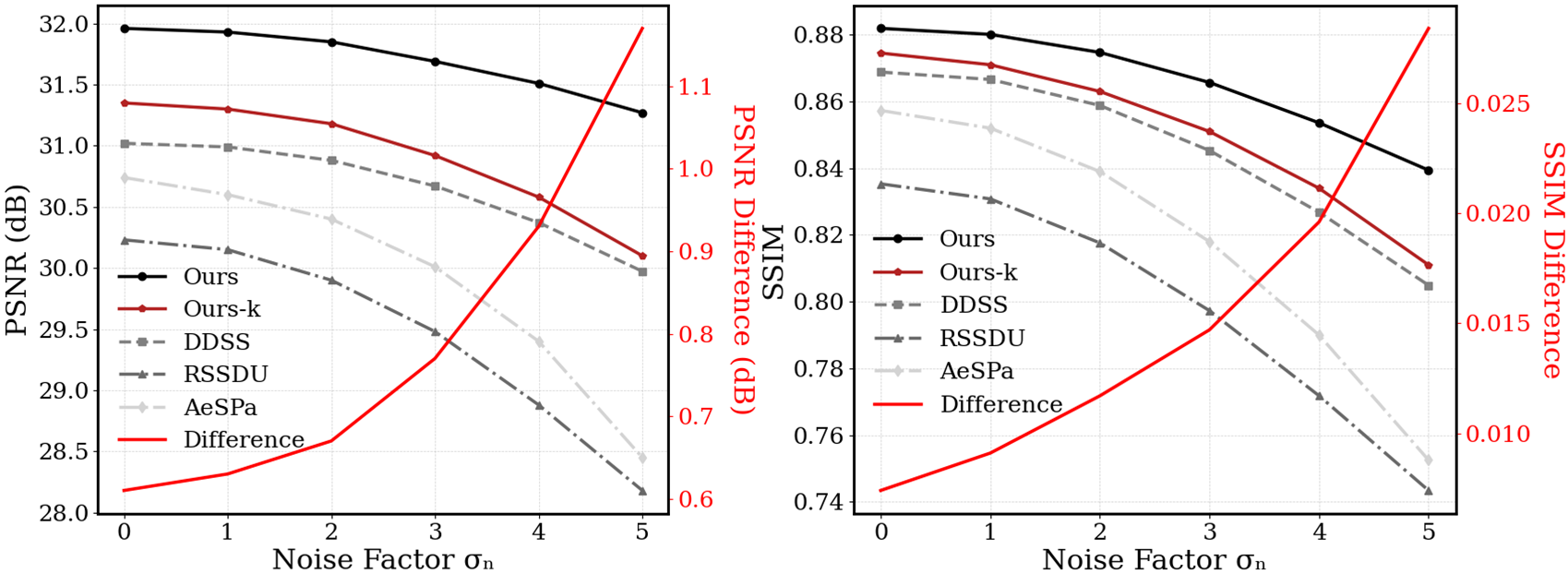} 
\caption{{Quantitative comparison of self-supervised methods on the Muti-coil Brain fastMRI dataset with different scales of simulated noise. Left y-axes are for reconstruction performance metrics (PSNR, SSIM) while the right y-axes are for
the “Difference” between Our proposed Method (the best) and the second best method.}}
\label{fig99}
\end{figure}

{In real medical imaging scenarios, the acquired k-space data is inevitably contaminated with noise. Therefore, evaluating the robustness of the proposed method against noise is crucial for its practical relevance in clinical applications. To this end, we design experiments to assess model performance under different noise levels. Specifically, all models are tested at 8$\times$ acceleration factors. In addition, we include a single-domain variant of our loss to investigate how the dual-domain loss interacts with noise. Inspired by \cite{b55}, noise was introduced by adding complex Gaussian white noise to the nonzero entries of the under-sampled k-space data during testing, simulating the effect of acquisition noise. The noise factor \( \sigma_n \) represents the relative noise amplitude (the ratio between the noise standard deviation and the signal magnitude) to avoid inconsistent noise scaling. It is worth noting that we removed the operator that directly replaces the reconstructed k-space data with the ground-truth k-space data at the sampled locations, since it is beneficial when the acquired data are clean but leads to performance degradation under noisy conditions.
As shown in Fig. \ref{fig99}, all methods suffer from performance degradation as the noise level increases. The red line indicates the performance gap between our model (the best) and the second-best one. It can be observed that this gap gradually widens with increasing noise level, suggesting that our method (with the dual-domain loss) exhibits the slowest performance degradation. This demonstrates the effectiveness of the dual-domain design and its superior generalization ability under noisy conditions.}

\subsection{Ablation Analysis}
\subsubsection{Effect of Number of Stages}
\begin{figure}[ht]
\centering 
\includegraphics[width=0.48\textwidth]{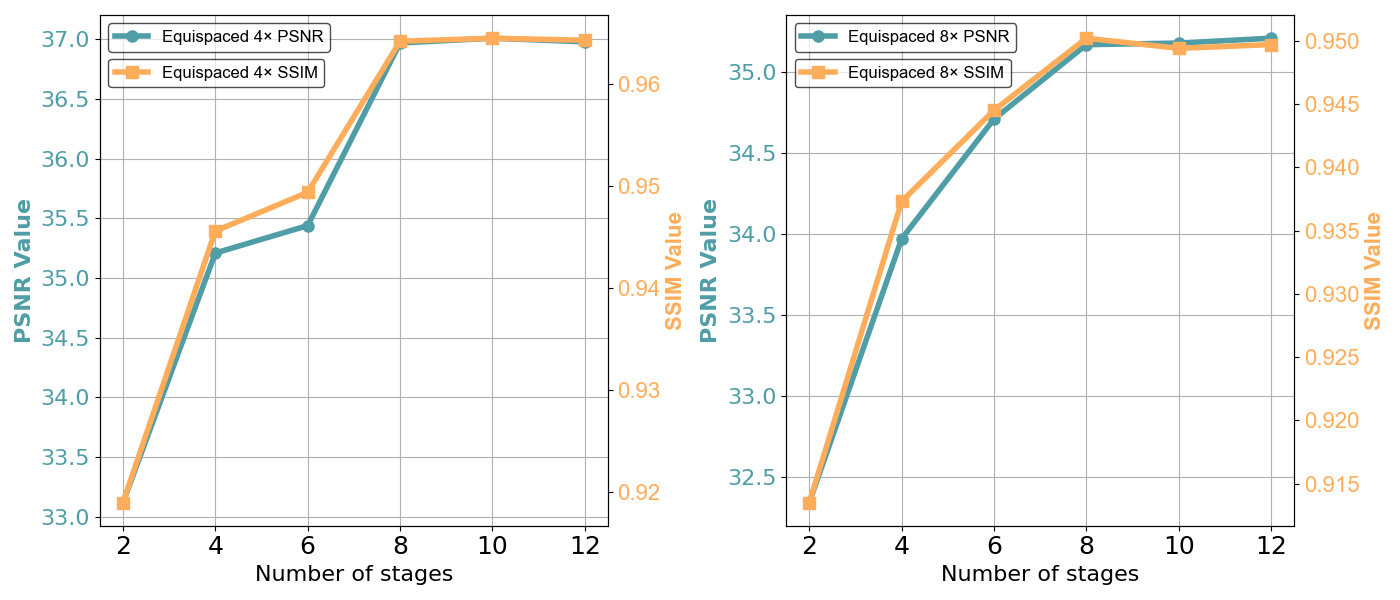} 
\caption{The PSNR and SSIM curves with different numbers of stages.}
\label{fig5}
\end{figure}

To illustrate how the number of stages $K$ influences reconstruction performance, we conduct a quantitative analysis of DUN-CP-PPA across different stage counts on four datasets, evaluating $4\times$ and $8\times$ acceleration using equispaced and random subsampling masks. As depicted in Fig. \ref{fig5}, the mean SSIM and PSNR values improve implicitly as $K$ increases from 2 to 8. However, after the eighth iteration, the improvements become marginal or even begin to decline. Considering both reconstruction performance and model complexity, we selected the model with $K=8$.

\subsubsection{Effectiveness of DUN-CP-PPA}
To evaluate the effectiveness of DUN-CP-PPA, we first replace the backbone under the same self-supervised learning framework (Ours). 
{We select HQS-Unet, ADMM-CSNet, and VarNet as backbones for comparison, as these models are widely used in MRI reconstruction tasks. Additionally, to verify the effectiveness of the SFFE block proposed in DUN-CP-PPA, we compare two variants of DUN-CP-PPA: $\text{DUN-CP-PPA}_S$, which captures only local features in the spatial domain, and $\text{DUN-CP-PPA}_F$, which captures only global features in the frequency domain. Notably, to ensure a fair comparison across different unfolding networks, namely HQS-Unet, ADMM-CSNet, and $\text{DUN-CP-PPA}_S$, we employed an identical feature extraction network for these three models. As shown in Table \ref{tb5}, the comparison among HQS-Unet, ADMM-CSNet, and $\text{DUN-CP-PPA}_S$ demonstrates the superiority of the proposed CP-PPA algorithm when integrated into a deep unfolding framework for MRI reconstruction. Regarding feature extraction strategies, the dual-domain version of DUN-CP-PPA achieves the best performance in all cases, further confirming the effectiveness of the proposed SFFE block.}

\begin{table}[!ht]
    \centering
    \scriptsize
    \renewcommand{\arraystretch}{0.96}
    \setlength\tabcolsep{8pt}
    \caption{Comparison with different network architectures under 4$\times$ and 8$\times$ acceleration in the re-visible dual-domain self-supervised learning approach.}
    \label{tb5}
    \begin{tabular}{ccc}
        \toprule
        \textbf{BackBone} & \textbf{4$\times$ Acceleration} & \textbf{8$\times$ Acceleration} \\
        \midrule
        \multirow{2}{*}{Varnet} & 34.53 $\pm$ 1.42 & 33.96 $\pm$ 1.49 \\
        & 0.9461 $\pm$ 0.0123 & 0.9365 $\pm$ 0.0119 \\
        \multirow{2}{*}{HQS-Unet} & 36.05 $\pm$ 1.37 & 34.46 $\pm$ 1.45 \\
        & 0.9571 $\pm$ 0.0085 & 0.9413 $\pm$ 0.0110 \\
        \multirow{2}{*}{{ADMM-CSNet}} & {36.43 $\pm$ 1.35} & {34.73 $\pm$ 1.48} \\
        & {0.9602 $\pm$ 0.0079} & {0.9458 $\pm$ 0.0106} \\
        \multirow{2}{*}{{$\text{DUN-CP-PPA}_S$}} & {36.77 $\pm$ 1.34} & {34.95 $\pm$ 1.51} \\
        & {0.9629 $\pm$ 0.0072} & {0.9487 $\pm$ 0.0103} \\
        \multirow{2}{*}{{$\text{DUN-CP-PPA}_{F}$}} & {36.27 $\pm$ 1.36} & {34.27 $\pm$ 1.45} \\
        & {0.9588 $\pm$ 0.0081} & {0.9398 $\pm$ 0.0112} \\
        \multirow{2}{*}{DUN-CP-PPA} & \textbf{36.97$\pm$ 1.32} & \textbf{35.17 $\pm$ 1.48} \\
        & \textbf{0.9642 $\pm$ 0.0070} & \textbf{0.9502 $\pm$ 0.0102} \\
        \bottomrule
    \end{tabular}
\end{table}

\subsubsection{Effectiveness of Loss Function}
{To demonstrate the effectiveness of our designed re-visible dual-domain self-supervised learning approach, we conduct ablation experiments by adding the loss terms one by one. As shown in Table \ref{tb6}, the overall experimental results demonstrate that the inclusion of additional loss terms leads to a gradual improvement in reconstruction metrics, indicating enhanced reconstruction performance. To verify the statistical significance of these improvements, each p-value was obtained from a paired Wilcoxon signed-rank test comparing the current setup with the previous one. It is noteworthy that the proposed $\mathcal{L}_{k}^{rev}$ contributes to a significant improvement in performance, while the dual-domain design of the loss function further boosts the reconstruction metrics. The best reconstruction performance is achieved when all the loss functions are jointly employed, which demonstrates the effectiveness of each proposed loss term.}

\begin{table}[!ht]
    \centering
    \scriptsize
    \renewcommand{\arraystretch}{0.96}
    \setlength\tabcolsep{4pt}
    \caption{Performance comparison of different loss functions. {Each p-value results from a paired Wilcoxon signed-rank test between the current and the previous setup.}}
    \label{tb6}
    \begin{tabular}{ccccccc}
        \toprule
        {\textbf{$\mathcal{L}_{k}^{reg}$}} & {\textbf{$\mathcal{L}_{k}^{rev}$}} & {\textbf{$\mathcal{L}_{\text{img}}^{reg}$}} & {\textbf{$\mathcal{L}_{\text{img}}^{rev}$}} & \textbf{4$\times$ Acceleration} & \textbf{8$\times$ Acceleration} & {\textbf{p-value}} \\
        \midrule
        {\multirow{2}{*}{\checkmark}} & {} & {} & {} & {34.72 $\pm$ 1.53} & {33.66 $\pm$ 1.33} & {\multirow{2}{*}{-}} \\
        {} & {} & {} & {} & {0.9425 $\pm$ 0.086} & {0.8879 $\pm$ 0.0194} & {} \\
        \multirow{2}{*}{{\checkmark}} & \multirow{2}{*}{{\checkmark}} & {} & {} & 35.40 $\pm$ 1.55 & 34.07 $\pm$ 1.34 & \multirow{2}{*}{{$<0.01$}} \\
        & & & & 0.9485 $\pm$ 0.0104 & 0.8948 $\pm$ 0.0188 & \\
        {\multirow{2}{*}{\checkmark}} & {\multirow{2}{*}{\checkmark}} & {\multirow{2}{*}{\checkmark}} & {} & {35.89 $\pm$ 1.32} & {34.70 $\pm$ 1.53} & {\multirow{2}{*}{$<0.01$}} \\
        {} & {} & {} & {} & {0.9605 $\pm$ 0.0076} & {0.9478 $\pm$ 0.0102} & {} \\
        \multirow{2}{*}{{\checkmark}} & \multirow{2}{*}{{\checkmark}} & \multirow{2}{*}{{\checkmark}} & \multirow{2}{*}{{\checkmark}} & \textbf{36.97$\pm$ 1.32} & \textbf{35.17 $\pm$ 1.48} & \multirow{2}{*}{{$<0.01$}} \\
        & & & & \textbf{0.9642 $\pm$ 0.0070} & \textbf{0.9502 $\pm$ 0.0102} & \\
        \bottomrule
    \end{tabular}
\end{table}

\subsubsection{Evaluation of Hyper-Parameter}
\begin{figure}[ht]
\centering 
\includegraphics[width=0.48\textwidth]{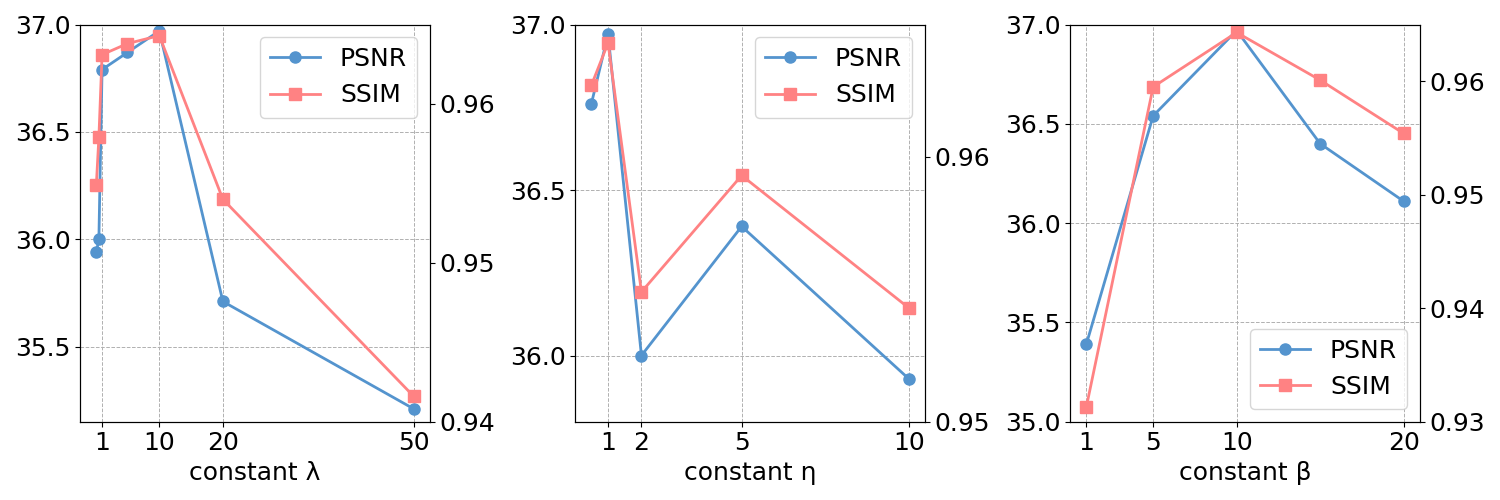} 
\caption{Impact of hyper-parameters $\lambda$, $\eta$, and $\beta$ on PSNR and SSIM. The figure shows how different values of these hyper-parameters affect the quality of reconstructed images, with metrics including PSNR (in blue) and SSIM (in red).}
\label{fig6}
\end{figure}

In this section, we evaluate the impact of hyperparameters in the loss function on the reconstruction performance and present the results in Fig. \ref{fig6}. Specifically, the hyperparameter $\beta$ is used to balance the loss functions in both the image domain and k-space, while $\eta$ controls the balance between the re-visible and regularization terms, which helps to stabilize the training process. The hyperparameter $\lambda$ {controls} the contribution of all k-space data as input during training. As shown in the figure \ref{fig6}, the different values of these hyperparameters significantly affect the quality of the reconstructed images, as measured by the PSNR and SSIM. We set \( \lambda \in \{0.1, 0.5, 1, 10, 20, 50\} \); \( \eta \in \{1, 2, 5, 10\} \); \( \beta \in \{1, 5, 10, 20\} \) for the experiments. Based on the observed trends, we select $\beta = 10$, $\eta = 1$, and $\lambda = 10$ as the optimal hyperparameter values to achieve the best reconstruction results.

\subsubsection{Model Verification}
\begin{figure}[!ht]
\centering 
\includegraphics[width=0.48\textwidth]{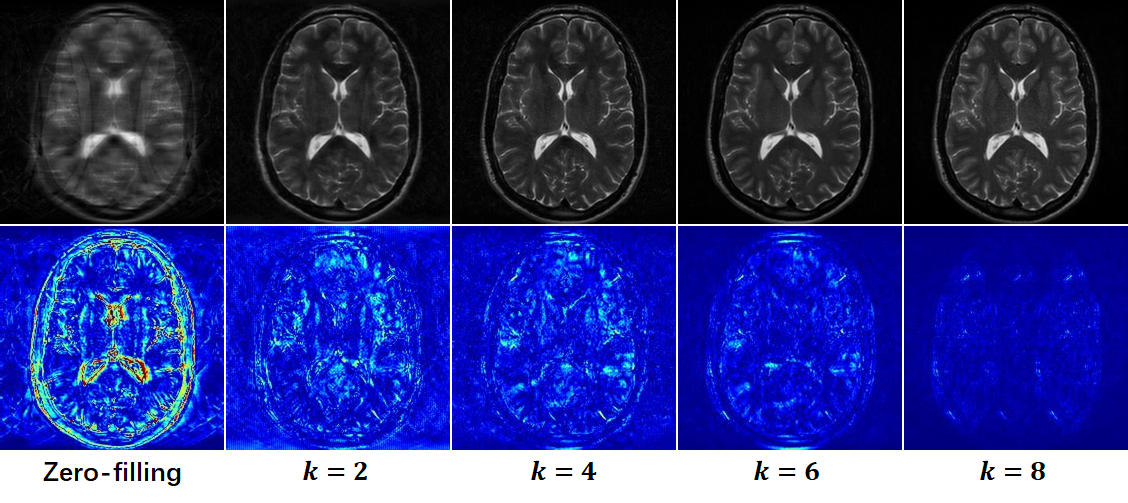} 
\caption{Visualization of reconstruction results at different stages \(k\).}
\label{fig7}
\end{figure}

In this section, we conduct a model verification experiment to demonstrate the reconstruction results of the proposed DUN-CP-PPA under different stages. As shown in Fig. \ref{fig7}, it is evident that as the number of stages \(k\) increases, the image quality improves. These results validate the design of our optimization-inspired iterative network, and the re-visible dual-domain self-supervised learning framework enables the proposed DUN-CP-PPA to achieve the expected MRI reconstruction. Furthermore, our network demonstrates superior transparency compared to other methods.

\section{Conclusion}
This paper proposes a re-visible dual-domain self-supervised deep unfolding network to imrove reconstruction performance using only under-sampled k-space data for training. Specifically, we design a re-visible dual-domain self-supervised learning framework and introduce a re-visible dual-domain loss function to enable the model to efficiently adapt to all under-sampled k-space data and {mitigate the input distribution shift between the the training stage and the inference stage. Furthermore, by unfolding each stage of CP-PPA into dedicated network modules, including SFFE block that jointly captures global and local representations, we embed imaging physics and comprehensive image priors to guide the reconstruction process, resulting in DUN-CP-PPA. Extensive experiments on both single-coil and multi-coil datasets demonstrate that our method outperforms state-of-the-art approaches in terms of reconstruction performance and generalization capability.}

\appendices
\section{Convergence discussion of DUN-CP-PPA}
\label{A1}
{
In reference \cite{b24}, a convergence proof is provided for the case where 
\(\mathrm{Prox}^{\tau_{k+1}}_{f}(x_k - \tau_{k+1}F_m^H y_k)\) 
is a special proximal operator for the \(\ell_1\)-norm, that is,
\(x_{k+1} = \widetilde{\mathcal{D}}\!\left(S_{\lambda_{k+1}}\!\big(\mathcal{D}(x_k - \tau_{k+1}F_m^H y_k)\big)\right)\).
Specifically, within the framework of MRI reconstruction, the proof is as follows.

With extrapolation \(\theta = 1\), we next conduct a convergence analysis of the iterative
framework of that special case of the DUN-CP-PPA framework.

First, consider that \(\mathcal{D}(x) = C_2 \mathrm{ReLU}(C_1 x)\),
where \(C_1\) and \(C_2\) are convolutional operators.
Let \(r_{k+1} = x_k - \tau_{k+1}F_m^H y_k\).
By approximating based on Theorem 1 \cite{b64}, we obtain:
\(\|\mathcal{D}(x) - \mathcal{D}(r_{k+1})\|_2^2 \approx \alpha \|x - r_{k+1}\|_2^2\),
where \(\alpha\) is a constant related to \(\mathcal{D}(\cdot)\) and can be merged into the learnable parameter \(\lambda_{k+1}\). Incorporating this relationship into 
\(x_{k+1} = \mathrm{Prox}^{\tau_{k+1}}_{f}(x_k - \tau_{k+1}F_m^H y_k) = \mathrm{Prox}^{\tau_{k+1}}_{f}(r_{k+1})\),
we derive:
\(x_{k+1} = \arg\min_{x} \tau_{k+1}\|\mathcal{D}(x)\|_{\ell_1} + \frac{1}{2}\|\mathcal{D}(x) - \mathcal{D}(r_{k+1})\|_2^2.\)
Thus, we obtain the closed form of \(\mathcal{D}(x_{k+1})\):
\(\mathcal{D}(x_{k+1}) = S_{\lambda_{k+1}}\!\big(\mathcal{D}(r_{k+1})\big).\)
Let \(\widetilde{\mathcal{D}}(\cdot)\) be the left inverse of \(\mathcal{D}(\cdot)\), we derive:
\(x_{k+1} = \widetilde{\mathcal{D}}\!\left(S_{\lambda_{k+1}}\!\big(\mathcal{D}(r_{k+1})\big)\right).\)
Thus, the optimal solution to 
\(x_{k+1} = \arg\min_{x} f(x) + \langle Lx, y_k\rangle + \frac{1}{2\tau_{k+1}}\|x - x_k\|^2\)
is proven to be
\(x_{k+1} = \widetilde{\mathcal{D}}\!\left(S_{\lambda_{k+1}}\!\big(\mathcal{D}(x_k - \tau_{k+1}F_m^H y_k)\big)\right).\)

\textbf{Lemma 1.}  
Let \(\chi \subset \mathbb{R}^n\) be a closed convex set, 
\(\xi(x)\) and \(\varphi(x)\) be convex functions, with \(\varphi(x)\) being differentiable. 
Assume that the solution set of the minimization problem 
\(\min_{x \in \chi} \{\xi(x) + \varphi(x)\}\) is nonempty. Then,
\(x^* \in \arg\min_{x \in \chi} \{\xi(x) + \varphi(x)\},\)
if and only if
\(x^* \in \chi,
\xi(x) - \xi(x^*) + (x - x^*)^T \nabla \varphi(x^*) \ge 0,  \forall x \in \chi.\)

Assuming that the pair \((x_{k+1}, y_{k+1})\) represents the optimal solution to Eq. \eqref{eq9},
we can obtain the following inequality according to Lemma1:
\[\left\{
\begin{aligned}
& f(x) - f(x_{k+1}) 
  + (x - x_{k+1})^{\mathrm{T}}\!\left(F_m^{H}y_k + \right.\\
& \left. \quad\frac{1}{\tau_{k+1}}(x_{k+1} - x_k)\right) \ge 0, \\[6pt]
& g^*(y) - g^*(y_{k+1}) 
  + (y - y_{k+1})^{\mathrm{T}}\!\left(-F_m z_{k+1} + \right.\\
& \left. \quad\frac{1}{\sigma_{k+1}}(y_{k+1} - y_k)\right) \ge 0.
\end{aligned} \tag{10} \label{eq10}
\right.\]

By defining \(u = (x, y)^{\mathrm{T}}\) and \(\zeta(u) = f(x) + g^*(y)\), we rewrite Eq. \eqref{eq10} as follows:
\[
\begin{aligned}
\zeta(u) - \zeta(u_{k+1}) 
+ (u - u_{k+1})^{\mathrm{T}}
\left(
\begin{pmatrix}
F_m^{H}y_k \\
- F_m z_{k+1}
\end{pmatrix}
\right. \\ 
\left.
+
\begin{pmatrix}
\frac{1}{\tau_{k+1}}(x_{k+1} - x_k) \\
\frac{1}{\sigma_{k+1}}(y_{k+1} - y_k)
\end{pmatrix}
\right)
\ge 0.
\end{aligned}
\]

From \(\theta = 1\), \(z_{k+1} = 2x_{k+1} - x_k\), we obtain:
\[
\begin{aligned}
\zeta(u) - \zeta(u_{k+1}) 
+ (u - u_{k+1})^{\mathrm{T}}
\!\left(
\begin{pmatrix}
F_m^{H}y_{k+1} \\
- F_m x_{k+1}
\end{pmatrix}
\right. \\ 
\left.
+
\begin{pmatrix}
\frac{1}{\tau_{k+1}}(x_{k+1} - x_k) - F_m^{H}(y_{k+1} - y_k) \\
- F_m (x_{k+1} - x_k) + \frac{1}{\sigma_{k+1}}(y_{k+1} - y_k)
\end{pmatrix}
\right)
\ge 0.
\end{aligned} \tag{11} \label{eq11}
\]

Hence, we can reformulate Eq. \eqref{eq11} as follows:
\[\zeta(u) - \zeta(u_{k+1}) + (u - u_{k+1})^{\mathrm{T}}\!\{G(u_{k+1}) + Q(u_{k+1} - u_k)\} \ge 0,\tag{12} \label{eq12}\]
where
\[
G = 
\begin{pmatrix}
0 & F_m^{H} \\[3pt]
- F_m & 0
\end{pmatrix},
\quad
Q = 
\begin{pmatrix}
\frac{1}{\tau_{k+1}} I & - F_m^{H} \\[3pt]
- F_m & \frac{1}{\sigma_{k+1}} I
\end{pmatrix}.
\]

Note that Eq. \eqref{eq12} is the generalized form of PPA \cite{b28,b65}, and in the case of 
\(\frac{1}{\tau_{k+1}\sigma_{k+1}} > \|F_m^{H}F_m\|_{\ell_2}\),
\(Q\) is symmetric and positive definite. Thus, the sequence \(\{x_k\}\) and \(\{y_k\}\) converge to the solution point of 
\(\min_x \max_y f(x) + \langle Lx, y\rangle - g^*(y).\)

Through the above proof, we can conclude that if the neural network \(\mathrm{ProxNet}_x\) 
can replace the proximal mapping \(\mathrm{Prox}^{\tau_{k+1}}_f\)
to efficiently compute the optimal solution of Eq. \eqref{eq9},
then the convergence proof of DUN-CP-PPA can be derived.
In our settings, we directly adopt a U-Net architecture as \(\mathrm{ProxNet}_x\),
enabling it to adaptively fit the appropriate \(\mathrm{Prox}^{\tau_{k+1}}_f\) under end-to-end training.

\section{Discussion about incorporating $\tilde{k}$ during training}
\label{A2}
\textbf{Mitigating covariate shift from training to inference:} Covariate shift occurs when the input distributions of the training and test data differ, while the conditional distributions of the labels given the inputs remain identical. In this setting, empirical risk minimization (ERM) may yield a model that is optimal under the training distribution but suboptimal under the test distribution.

Specifically, according to reference \cite{b66}: \(h_Q^* \in \arg\min_{h \in \mathcal{H}} L_Q(h, f_Q), \quad h_P^* \text{ be a minimizer of } L_P(h, f_P).\) These minimizers may not be unique, but for adaptation to succeed, it is natural to assume that the average loss 
\( L_Q(h_Q^*, h_P^*) \) between the best-in-class hypotheses is small. The following theorem provides a bound on the error of a hypothesis with respect to the target domain.

Assume that the loss function \( L \) is symmetric and satisfies the triangle inequality. Then, for any hypothesis \( h \in \mathcal{H} \), the following holds:
\begin{align*}
L_P(h, f_P) &\le L_P(h_P^*, f_P) + L_Q(h, h_Q^*) + \mathrm{disc}_L(P, Q) \\
&+ \min\{L_Q(h_Q^*, h_P^*), L_P(h_Q^*, h_P^*)\},
\end{align*}
where \( \mathrm{disc}_L(P, Q) \) denotes the discrepancy distance between the training input distribution and the test input distribution under the task-specific loss \( L \). The final term, 
\(\min\{L_Q(h_Q^*, h_P^*), L_P(h_Q^*, h_P^*)\}\), quantifies the inherent mismatch between the optimal hypotheses on the two domains. When the labeling functions are consistent across domains, this term becomes negligible (i.e., \( \approx 0 \)).

Under the covariate-shift assumption, the optimal hypotheses \( h_P^* \) and \( h_Q^* \) coincide, causing the last term to vanish. In this case, the bound simplifies to:
\[
L_P(h, f_P) \le L_P(h_P^*, f_P) + L_Q(h, h_Q^*) + \mathrm{disc}_L(P, Q),
\]
indicating that the generalization gap primarily depends on the discrepancy between the training input distribution and the test input distribution. When this divergence is large, models trained via ERM may achieve low source-domain loss yet perform suboptimally on the target domain.

To mitigate this issue, additional information available at inference time (e.g., $\tilde{k}$) should be incorporated during training through specially design. In our method, by explicitly incorporating $\tilde{k}$ as an input during training, the model can more easily leverage features corresponding to $\tilde{k}$ at inference. Since $\tilde{k}$ participates in the forward pass and influences the gradients through the loss function, the model is encouraged to optimize its performance with respect to $\tilde{k}$, rather than focusing solely on exploiting $\tilde{k}_p$.}

\end{document}